\documentclass[aps,prb,preprint,floatfix,superscriptaddress]{revtex4-2}

\usepackage{graphicx}% Include figure files
\usepackage{dcolumn}% Align table columns on decimal point
\usepackage{bm}% bold math
\usepackage{braket}
\usepackage{xcolor}
\usepackage{tikz}
\usepackage{ulem}
\usepackage[caption=false]{subfig}
\usepackage{mhchem}
\usetikzlibrary{quantikz2}
\usepackage{multirow}
\usepackage{makecell}
\usepackage{amsmath, amssymb, amsthm,mathtools,mathrsfs}
\usepackage[hmargin=1in,vmargin=1in]{geometry}
\usepackage{microtype}
\usepackage{enumitem}
\usepackage{soul}
\usepackage{float}
\usepackage{appendix}
\usepackage[table,xcdraw]{xcolor}
\usepackage[utf8]{inputenc}
\usepackage[T1]{fontenc}
\usepackage{textcomp}

\usepackage{rotating}
\usepackage[colorlinks=true, linkcolor=blue, citecolor=blue, urlcolor=blue]{hyperref}

\preprint{APS/123-QED}

\begin{document}
\sloppy
\newcommand{\bea}{\begin{eqnarray}}
\newcommand{\eea}{\end{eqnarray}}
\newcommand{\EQ}[1]{Equation~(\ref{#1})} %
\newcommand{\eq}[1]{Eq.~(\ref{#1})} %
\newcommand{\eqs}[1]{Eqs.~(\ref{#1})} %
\newcommand{\fig}[1]{Fig.~\ref{#1}} %
\newcommand{\figs}[1]{Figs.~\ref{#1}} %
\newcommand{\CR}[1]{\hat a^{\dagger}_{#1}}
\newcommand{\AN}[1]{\hat a_{#1}}

\newcommand{\g}{\mathfrak g}
\newcommand{\rfrak}{\mathfrak r}
\newcommand{\sfrak}{\mathfrak s}
\newcommand{\Z}{\mathrm{Z}}
\newcommand{\ad}{\mathrm{ad}}
\newcommand{\Span}{\mathrm{span}}
\newcommand{\br}[2]{\left[ #1, #2 \right]}

% Toggle for marked-up (blue) vs clean (black) version:
\newcommand{\BC}[1]{\textcolor{black}{#1}}

\title{Exact Factorization of Unitary Transformations with Spin-Adapted Generators}

\author{Paarth Jain}
\affiliation{Department of Physics, University of Toronto, Toronto, Ontario M5S 3H6, Canada}
\affiliation{Chemical Physics Theory Group, Department of Chemistry, University of Toronto, Toronto, Ontario M5S 3H6, Canada}

\author{Artur F. Izmaylov}
\email{artur.izmaylov@utoronto.ca}
\affiliation{Chemical Physics Theory Group, Department of Chemistry, University of Toronto, Toronto, Ontario M5S 3H6, Canada}
\affiliation{Department of Physical and Environmental Sciences, University of Toronto Scarborough, Toronto, Ontario M1C 1A4, Canada}

\author{Erik R. Kjellgren}
\email{kjellgren@sdu.dk}
\affiliation{Department of Physics, Chemistry and Pharmacy,
University of Southern Denmark, Campusvej 55, 5230 Odense, Denmark.}
\date{\today}

\begin{abstract}
Preserving spin symmetry in variational quantum algorithms is essential for producing physically meaningful electronic wavefunctions. Implementing spin-adapted transformations on quantum hardware, however, is challenging because the corresponding fermionic generators translate into noncommuting Pauli operators.
In this work, we introduce an exact and computationally efficient factorization of spin-adapted unitaries derived from fermionic double excitation and deexcitation rotations. These unitaries are expressed as ordered products of exponentials of Pauli operators. Our method exploits the fact that the elementary operators in these generators form small Lie algebras. By working in the adjoint representation of these algebras, we reformulate the factorization problem as a low-dimensional nonlinear optimization over matrix exponentials. This approach enables precise numerical reparametrization of the unitaries without relying on symbolic manipulations.
The proposed factorization provides a practical strategy for constructing symmetry-conserving quantum circuits within variational algorithms. It preserves spin symmetry by design, reduces implementation cost, and ensures the accurate representation of electronic states in quantum simulations of molecular systems.
\end{abstract}

\maketitle

\section{Introduction}

Preserving spin and particle-number symmetries in variational quantum algorithms (VQAs) is essential for accurately describing molecular electronic structure. These symmetries ensure physically consistent potential energy surfaces, accelerate convergence, and prevent artificial symmetry breaking. However, maintaining them within scalable ansätze remains a significant challenge for quantum simulations of electronic systems.

Conventional unitary coupled cluster (UCC) ansätze\cite{UCC:rev} and their adaptive extensions\cite{QCC,iQCC,Grimsley2019-zw,Qubit-ADAPT,Burton2023} construct the variational wavefunction from a pool of anti-Hermitian generators chosen to balance expressiveness and circuit depth. In principle, the complete set of $4^N - 1$ traceless Pauli generators acting on $N$ qubits spans the full $\mathfrak{su}(2^N)$ Lie algebra, guaranteeing convergence to the exact ground state within a fixed orbital basis. The exponential scaling of this set, however, makes its direct use computationally infeasible. To achieve polynomial scaling, smaller operator pools that are closed under commutation have been proposed in both fermionic\cite{Grimsley2019-zw} and qubit operator spaces\cite{Qubit-ADAPT}. While such reduced pools can in principle generate the full algebra, many of them break key physical symmetries, including total spin ($S^2$), its projection ($S_z$), and particle number ($N_e$).

Maintaining these symmetries during variational optimization is desirable not only for physical consistency but also for improved convergence and reduced redundancy in the variational parameters.
\BC{Indeed, symmetry breaking in operator pools has been shown to slow the convergence of adaptive VQE algorithms~\cite{Bertels2022}.}
Several strategies have been proposed to enforce or restore symmetry, including penalty-based constraints\cite{ryabinkin2019constr}, projection methods\cite{ProjEA2019}, and modified symmetry-conserving generators\cite{Anselmetti_2021,Burton2023}. The DISCO framework\cite{Burton2023}, for example, introduces one-electron and electron-pair rotations that preserve total spin by construction. Although effective for spin conservation, this approach restricts the operator pool to zero-seniority excitations (pair rotations), thereby reducing the overall expressivity of the ansatz.

Achieving full spin conservation while retaining the expressivity associated with all symmetrized double excitation and deexcitation rotations remains an open challenge. Implementing such transformations on quantum hardware requires their factorization into sequences of unitaries that can be compiled into standard gate sets. In this work, we address this gap by developing a numerically efficient and exact factorization scheme for unitaries generated by spin-adapted double excitation and deexcitation operators.

The primary difficulty in implementing these spin-adapted transformations arises because the corresponding generators are linear combinations of noncommuting operators in both fermionic and qubit representations. Standard Trotter-based decompositions therefore produce only approximate unitaries. The key insight of this work is that double excitation and deexcitation fermionic operators form compact Lie algebras. By exploiting the Lie algebraic closure, we rewrite the exponential of a sum of elementary fermionic generators as an ordered product of exponentials of the algebra’s basis elements, each of which can be implemented directly as a quantum circuit.

Since the number of basis elements in the relevant Lie algebras is modest (fewer than 100), the entire factorization can be carried out efficiently in the adjoint representation using low-dimensional matrix algebra. Instead of employing symbolic Baker–Campbell–Hausdorff expansions, we reduce the reparametrization problem to a small nonlinear optimization involving matrix exponentials in the adjoint representation. This numerical treatment achieves high precision while avoiding symbolic complexity.

\BC{Although we benchmark the factorization within variational quantum eigensolver (VQE) calculations for concreteness, the decomposition itself is algorithm-agnostic. It depends only on the Lie algebra structure of the generators, not on how the rotation parameters are determined. In principle, the same algebraic framework could be adapted to Trotterized Hamiltonian simulation. However, because the electronic Hamiltonian is Hermitian rather than anti-Hermitian, the spin-conserving partitioning and the resulting Lie algebras would differ from those treated here, and a separate assessment of accuracy relative to existing Trotterization schemes would be needed. The factorized unitaries are also useful as compact initial-state preparations for quantum phase estimation (QPE) or other fault-tolerant algorithms.}

The remainder of this paper presents the theoretical framework for this reparametrization, analyzes the structure of the relevant Lie algebras for different seniority sectors, and demonstrates numerical factorization for representative spin-adapted double excitation and deexcitation transformations. We conclude by discussing the implications of this factorization for constructing compact, symmetry-preserving variational ans\"atze in quantum chemistry.

\BC{While completing this work, we became aware of the
work of Magoulas and Evangelista\cite{Magoulas2025-bz}, who has also derived exact circuit representations for the spin-adapted double excitation operators.}

\section{Method}

\subsection{Lie algebra framework}

The goal of this work is to express unitaries that cannot be directly compiled into gate sequences as products of unitaries whose gate decompositions are known. 
Unitaries with spin-symmetrized generators can be written as
\[
\exp{\left[\sum_k c_k A_k\right]},
\]
where $A_k \in \mathcal{A}$ form a Lie algebra generated by the commutator closure of the $\{A_k\}$ operators. 
The dimensionality of these algebras, $m \lesssim 84$, depends on the seniority and the specific index pattern. 
Each separate exponential $\exp(t A_k)$ admits a well-known gate representation, and therefore our objective is to decompose
\begin{align}\label{eq:decomp}
 e^{\theta (\sum_k c_k A_k)} &= \prod_p e^{t_p(\theta) A_p}.
\end{align}

The first question is whether such a product representation exists for the Lie algebras considered. 
This issue is nontrivial because certain Lie algebras contain generator sets for which the exponential of a linear combination cannot be represented as a finite product of exponentials of the individual generators~\cite{nonSurj}. 
In our case, all generators are anti-Hermitian operators that form subalgebras of the compact $\mathfrak{su}(N)$ algebra, for which the exponential map is surjective. 
Consequently, any element of the Lie group generated by the left-hand side (LHS) of Eq.~\eqref{eq:decomp} can be expressed by an appropriate choice of parameters in the right-hand side (RHS).

To perform the decomposition, we employ a low-dimensional faithful matrix representation of the Lie algebras rather than symbolic Baker--Campbell--Hausdorff (BCH) expansions. 
This converts the factorization into a numerical matrix optimization problem, solved by minimizing the norm of the difference between the LHS and RHS of Eq.~\eqref{eq:decomp}. 
A well-known alternative is the Wei--Norman reparametrization approach, which leads to a system of differential equations whose number equals the dimensionality of the Lie algebra. 
In our case, the algebras are large enough that analytical solutions are impractical, making a numerical approach more efficient. 
Moreover, the Wei--Norman method requires continuous trajectories $t_p(\theta)$ from $\theta = 0$ to the target value, which may impose additional constraints on the Lie algebra beyond the surjectivity of the exponential map.

\subsection{Lie algebras}

We consider three Lie algebras corresponding to the following cases: seniority-2, seniority-4 intermediate singlet, and seniority-4 intermediate triplet. 
These algebras arise from the commutator closures of two-electron rotation operators $\hat{G}_{i_\sigma j_\tau}^{a_\mu b_\nu}$ present in every symmetrized generator:
\begin{equation}
    \hat{G}_{i_\sigma j_\tau}^{a_\mu b_\nu} =
    \hat{a}_{a_\mu}^\dagger \hat{a}_{b_\nu}^\dagger
    \hat{a}_{j_\tau} \hat{a}_{i_\sigma}
    - \hat{a}_{i_\sigma}^\dagger \hat{a}_{j_\tau}^\dagger
    \hat{a}_{b_\nu} \hat{a}_{a_\mu},
\end{equation}
where $i,j,a,b$ are spatial orbital indices and $\sigma,\tau,\mu,\nu$ are spin projection indices.
The corresponding unitary transformations are generated by symmetrized combinations of double excitations and deexcitations.

\textit{Seniority-2:}
\bea
\hat G^{\mathrm{SA}}_{iiab}
&=&\frac{1}{\sqrt{2}}\Big(\hat G^{a_\alpha b_\beta}_{i_\alpha i_\beta}
-\hat G^{a_\beta b_\alpha}_{i_\alpha i_\beta}\Big),
\label{eq:2s}\\
\hat G^{\mathrm{SA}}_{ijaa}
&=&\frac{1}{\sqrt{2}}\Big(\hat G^{a_\alpha a_\beta}_{i_\alpha j_\beta}
-\hat G^{a_\alpha a_\beta}_{i_\beta j_\alpha}\Big).
\eea

\textit{Seniority-4, intermediate singlet:}
\begin{equation}\label{eq:4s}
\hat G^{\mathrm{SA}}_{ijab}
=\frac{1}{2}\Big(
\hat G^{a_\alpha b_\beta}_{i_\alpha j_\beta}
+\hat G^{a_\beta b_\alpha}_{i_\beta j_\alpha}
-\hat G^{a_\alpha b_\beta}_{i_\beta j_\alpha}
-\hat G^{a_\beta b_\alpha}_{i_\alpha j_\beta}
\Big),
\end{equation}

\textit{Seniority-4, intermediate triplet:}
\begin{align}\label{eq:4t}
\hat G^{\mathrm{SA}'}_{ijab}
&=\frac{1}{2\sqrt{3}}\Big(
\hat G^{a_\alpha b_\beta}_{i_\alpha j_\beta}
+\hat G^{a_\beta b_\alpha}_{i_\beta j_\alpha}
+\hat G^{a_\alpha b_\beta}_{i_\beta j_\alpha}
+\hat G^{a_\beta b_\alpha}_{i_\alpha j_\beta}
\nonumber\\
&\qquad\qquad
+2\,\hat G^{a_\alpha b_\alpha}_{i_\alpha j_\alpha}
+2\,\hat G^{a_\beta b_\beta}_{i_\beta j_\beta}
\Big).
\end{align}

\BC{The spin-adapted unitaries generated by Eqs.~\eqref{eq:2s}--\eqref{eq:4t} conserve total spin $\hat{S}^2$ and its projection $\hat{S}_z$ for any spin quantum number $S$, not only for singlet or triplet states. Each symmetrized generator commutes with both $\hat{S}^2$ and $\hat{S}_z$ by construction, so its exponential preserves the spin sector of any initial state. The labels intermediate singlet and intermediate triplet in Eqs.~\eqref{eq:4s} and~\eqref{eq:4t} refer to the intermediate coupling scheme used to construct the two-electron operators.}

We summarize the closures for all three cases, providing explicit expressions for the seniority-2 algebra in this section and the seniority-4 algebras in Appendices~\ref{app:sen4s}--\ref{app:sen4t}. 
All algebras share several structural features, which we illustrate using the seniority-2 example.

First, the commutator closure of generators of the form $\hat{G}_{i_\sigma j_\tau}^{a_\mu b_\nu}$ produces additional elements of the type $p[\{\hat n(q_\sigma)\}] \hat{G}_{i_\sigma j_\tau}^{a_\mu b_\nu}$,
where $p[\{\hat n(q_\sigma)\}]$ are polynomials of occupation number operators $\hat n(p_\sigma)=\hat a^\dagger_{p_\sigma}\hat a_{p_\sigma}$. 
These polynomials have only three eigenvalues $\{\pm 1, 0\}$. 
Their indices do not overlap with those in $\hat{G}_{i_\sigma j_\tau}^{a_\mu b_\nu}$, and the two factors in each element commute,
\([p[\{\hat n(q_\sigma)\}],\hat{G}_{i_\sigma j_\tau}^{a_\mu b_\nu}] = 0.\)
Because both factors have spectra $\{\pm 1,0\}$, the total spectrum of each Lie algebra generator is also $\{\pm 1,0\}$.

For example, one basis of the seniority-2 Lie algebra is
\begin{align}
A_1 &= \hat G^{a_\alpha b_\beta}_{i_\alpha i_\beta},\\
A_2 &= \hat G^{a_\beta b_\alpha}_{i_\alpha i_\beta},\\
A_3 &=(1-\hat n(i_\alpha)-\hat n(i_\beta))\,
          \hat G^{a_\beta b_\alpha}_{a_\alpha b_\beta},\\
A_4 &= -\Big(1-\hat n(a_\alpha)-\hat n(b_\beta)+2\hat n(a_\alpha)\hat n(b_\beta)\Big)\,
          \hat G^{a_\beta b_\alpha}_{i_\alpha i_\beta},\\
A_5 &= \Big(1-\hat n(a_\beta)-\hat n(b_\alpha)+2\hat n(a_\beta)\hat n(b_\alpha)\Big)\,
          \hat G^{a_\alpha b_\beta}_{i_\alpha i_\beta},
\end{align}
such that $\mathcal{A}_{S2} = \{A_i\}_{i=1}^5$. 
These basis elements satisfy the commutation relations
\begin{align*}
&[A_1,A_2]=A_3,\quad [A_1,A_3]=A_4,\quad [A_1,A_4]=-A_3,\quad [A_1,A_5]=0,\\
&[A_2,A_3]=A_5,\quad [A_2,A_4]=0,\quad [A_2,A_5]=-A_3,\\
&[A_3,A_4]=A_5,\quad [A_3,A_5]=-A_4,\quad [A_4,A_5]=A_3.
\end{align*}

Second, all algebras possess nontrivial centers $Z(\mathcal{A})$, defined as the set of elements that commute with all others. 
The dimensionality of each center equals the number of $\hat{G}_{i_\sigma j_\tau}^{a_\mu b_\nu}$ terms in the symmetrized generator (i.e., 2, 4, and 6). 
All central elements are of the form 
$P[\{\hat n(q_\sigma)\}] \hat{G}_{i_\sigma j_\tau}^{a_\mu b_\nu}$, 
where $P[\{\hat n(q_\sigma)\}]$ are projector polynomials of occupation numbers with eigenvalues $\{0,1\}$. 
In the seniority-2 case, there are two central elements, $Z_1 = A_1 - A_5$ and $Z_2 = A_2 + A_4$, such that 
$Z(\mathcal{A}_{S2}) = \mathrm{span}\{Z_1, Z_2\} \cong \mathbb{R}^2$. 
Their explicit forms are
\[
\begin{aligned}
Z_{1} &= \big(\hat{n}(j_\beta)-\hat{n}(a_\alpha)\big)^{2}\,\hat{G}_{i_\alpha i_\beta}^{\, j_\alpha a_\beta},\\[4pt]
Z_{2} &= \big(\hat{n}(j_\alpha)-\hat{n}(a_\beta)\big)^{2}\,\hat{G}_{i_\alpha i_\beta}^{\, j_\beta a_\alpha}.
\end{aligned}
\]

Finally, each Lie algebra can be partitioned as a direct sum of its center and a semisimple part, which itself decomposes into simple ideals. 
For the seniority-2 case, the subspace 
\(\mathfrak s := \mathrm{span}\{A_3, A_4, A_5\}\) 
is an ideal and is isomorphic to 
\(\mathfrak{so}(3)\cong \mathfrak s.\) 
Thus, the full Lie algebra can be expressed as 
\(\mathcal{A}_{S2} = \mathfrak s \oplus Z(\mathcal{A}_{S2}) \cong \mathfrak{so}(3)\oplus\mathbb{R}^2.\) 
Although this partitioning does not reduce the number of terms in Eq.~\eqref{eq:decomp}, it facilitates the construction of the decomposition in the adjoint representation.

\subsection{Adjoint representation}

To determine the parameters in Eq.~\eqref{eq:decomp} we use the adjoint representation, which maps each algebra element to a matrix of dimension equal to that of the Lie algebra. 
For the algebras considered here, the dimensions (depending on seniority and index pattern) are $5$, $28$, and $84$. 
These sizes are substantially smaller than those arising in faithful representations built in the computational basis of the corresponding spin orbitals. 
For example, using the computational basis in the Fock space of the involved spin orbitals requires matrices of dimensions $64$ and $256$ for seniorities $2$ and $4$, respectively. 
Therefore, we adopt the adjoint representation, in which operator matrices are constructed from the commutator (structure) constants: each element $A_i$ corresponds to a matrix $M^{(i)}$ defined by
\bea
[A_i,A_j] = \sum_k M_{jk}^{(i)} A_k,
\eea
so that the entries of $M^{(i)}$ are the structure constants of the Lie algebra.

This representation is faithful on the Lie algebra modulo its center: all central elements map to the zero matrix because they commute with every element. 
Since all algebras studied here have nontrivial centers, we separate the central and semisimple contributions in Eq.~\eqref{eq:decomp} before applying the adjoint representation. 
The additional commutativity within the center provides a universal reduction that isolates the semisimple component, allowing us to carry out the decomposition entirely in the adjoint representation of the semisimple part.

For all three algebras, the initial unitary can be written as
\bea
U(\theta) = \exp\left[\theta \sum_{i=1}^{K} c_i A_i \right],
\eea
where $c_i$ are fixed coefficients [see Eqs.~\eqref{eq:2s}, \eqref{eq:4s}, and \eqref{eq:4t}], and $K$ is the number of central elements in the algebra. 
Each central element can be written as $Z_i = A_i + S_i$, where $S_i = \sum_s d_{is} A_s$ involves only elements from the semisimple part. 
In general, the semisimple elements do not mutually commute, but all elements forming the $Z_i$ commute with one another. 
Using this commutativity, we obtain
\bea 
\exp\left[\theta \sum_{i=1}^{K} c_i A_i\right] &=& \exp\left[\theta \sum_{i=1}^{K} c_i (Z_i - S_i)\right] \\
&=&  \exp\left[\theta \sum_{i=1}^{K} c_i Z_i \right] \exp\left[-\theta \sum_{i=1}^{K} c_i S_i\right].
\eea
The central part then factors as
\bea\label{eq:decompc}
\exp\left[\theta \sum_{i=1}^{K} c_i Z_i \right] = \prod_i e^{\theta c_i Z_i},
\eea
and, since each $Z_i$ reduces to a linear combination of mutually commuting Pauli products, these exponentials are straightforward to implement. 
The combined semisimple contribution involves only elements $\{A_s\}$ from the semisimple part,
\bea\label{eq:decomps}
U_S(\theta) = \exp\left[-\theta \sum_{i=1}^{K} c_i S_i\right] 
= \prod_s e^{t_s(\theta) A_s} = U_{S2}(\mathbf t).
\eea
Because the adjoint representation is faithful on the semisimple subalgebra, we use it to fit the parameters $t_s(\theta)$ numerically.

We determine $t_s(\theta)$ by minimizing the Frobenius norm of the difference between the two sides of Eq.~\eqref{eq:decomps}, evaluated via matrix exponentials:
\begin{equation}\label{eq:cost}
\mathbf t^\star(\theta)=\arg\min_{\mathbf t\in\mathbb{R}^m}\;\big\| U_S(\theta)- U_{S2}(\mathbf t)\big\|_F^2.
\end{equation}
This objective is representation-invariant within $\exp(\mathfrak s)$ and encodes the Baker--Campbell--Hausdorff identity for the chosen semisimple Lie algebra. 
In practice, we employ a smooth quasi-Newton optimizer (BFGS). 
Because $m\le 24$ in all cases considered, each evaluation of the norm difference is inexpensive and convergence is rapid.
Code to perform this minimisation is available on GitHub\cite{spinadapted}.

\BC{The optimizer is initialized by expressing the semisimple component of the generator in the orthonormal basis $\{A_s\}$ of the semisimple subalgebra, and the resulting coordinates serve as the starting point $\mathbf{t}_0$. We use a gradient tolerance of $10^{-10}$ and a maximum of $500$ iterations, and in all cases tested the Frobenius-norm residual $\|U_S - U_{S2}\|_F$ falls below $10^{-6}$. Because the generators form subalgebras of the compact algebra $\mathfrak{su}(N)$, the exponential map is surjective, guaranteeing that a factorization $U_{S2}(\mathbf{t}^\star) = U_S(\theta)$ exists for every value of $\theta$. The factorization is therefore exact in the mathematical sense, as no Trotter-type approximation is involved, and the role of the optimizer is solely to determine the numerical values of the parameters $\mathbf{t}^\star$. For the seniority-4 intermediate triplet case, the semisimple part decomposes into $26$ mutually commuting $\mathfrak{so}(3)$ ideals (Appendix~\ref{app:sen4t}), each of which admits a closed-form $ZYZ$ Euler-angle decomposition $e^{x_1 B_1 + x_2 B_2 + x_3 B_3} = e^{\alpha B_3}\,e^{\beta B_2}\,e^{\gamma B_3}$, bypassing the numerical optimization entirely and yielding machine precision accuracy.}

\subsection{Implementation of unitaries}
\label{sec:unitaries-impl}

All semisimple generators ($A_s$) and central generators ($Z_i$) in Eqs.~\eqref{eq:decomps} and \eqref{eq:decompc} map, under a fermion-to-qubit transform, to linear combinations of commuting Pauli products within each factor. 
For definiteness, we use the Jordan--Wigner (JW) mapping. 
Given $\mathbf t(\theta)$, each factor $\exp\!\big(t_s(\theta) A_s\big)$ in Eq.~\eqref{eq:decomps} and each central factor $\exp(\theta c_i Z_i)$ in Eq.~\eqref{eq:decompc} can be expressed as a product of exponentials of individual Pauli products. 
Each commuting set is implemented by a standard sequence: basis change, CNOT ladder, $R_z$ rotation, and inverse ladder. 
Table~\ref{tab:num_pauli} reports Pauli-string counts under JW for the factorizations described.

\begin{table}[H]
    \centering
    \begin{tabular}{c|c|c|c|c}
         & $\hat{G}^\text{SA}_{iiab}$ & $\hat{G}^\text{SA}_{ijaa}$ & $\hat{G}^\text{SA}_{ijab}$ (S) & $\hat{G}^{\text{SA}'}_{ijab}$ (T)  \\
        \hline
        $N_\text{Pauli}^{\text{JW}}$ & 48 & 48 & 384 & 640
    \end{tabular}
    \caption{Number of Pauli strings in the factorized implementation under the JW mapping.}
    \label{tab:num_pauli}
\end{table}

To reduce the Pauli-string count, the generators ($A_s$) are grouped into groups of commuting operators.
For all of the three cases, seniority-2, seniority-4 (intermediate singlet), and, seniority-4 (intermediate triplet), this results in three groups.
The groups of commuting generators for the seniority-2 case are,

\begin{align}
s&=\{1,5\},\quad s=\{2,4\},\quad s=\{3\}
\end{align}

The groups for the seniority-4 cases can be found in Appendices~\ref{app:sen4s}--\ref{app:sen4t}.

\section{Results}

\subsection{Computational details}

To evaluate the effect of spin-adapted fermionic generators on the efficiency and accuracy of variational algorithms, we carried out a series of adaptive VQE calculations using different operator pools. 
All calculations employed the STO-3G\cite{Hehre1969-ju,Hehre1970-ym} basis set and were converged when the maximum absolute value of the gradient of the operator pool was below $10^{-5}$~a.u.
The state-vector simulator implemented in \textsc{SlowQuant}\cite{slowquant} was used to perform noiseless classical emulations. 
Energy minimizations employed the BFGS algorithm as implemented in \textsc{SciPy}\cite{2020SciPy-NMeth}.

The operator pool for the conventional fermionic singles and doubles (SD) adaptive fermionic ansatz consisted of 
$\{\hat{G}_{i_\sigma}^{a_\sigma},\,\hat{G}_{i_\sigma j_\tau}^{a_\sigma b_\tau},\,\hat{G}_{i_\sigma j_\tau}^{a_\tau b_\sigma}\},\quad i,j<a,b$, this ansatz is equal to the ADAPT ansatz\cite{Grimsley2019-zw}.
For the spin-adapted fermionic ansatz ($\mathrm{S}_\mathrm{SA}\mathrm{D}_\mathrm{SA}$), the operator pool was $\{\hat{G}^\mathrm{SA}_{ia},\,\hat{G}^\mathrm{SA}_{ijab},\,\hat{G}^{\mathrm{SA}'}_{ijab}\},\quad i,j<a,b$.
Here, $\hat{G}^\mathrm{SA}_{ia}=\hat{G}_{i_\alpha}^{a_\alpha} + \hat{G}_{i_\beta}^{a_\beta}$, is a fermionic spin-adapted single excitation operator.
Finally, for the spin-adapted singles and pair-doubles ansatz ($\mathrm{S}_\mathrm{SA}\mathrm{D}_\mathrm{pair}$), the operator pool was 
$\{\hat{G}^\mathrm{SA}_{ia},\,\hat{G}^\mathrm{SA}_{iiaa}\},\quad i<a,$
where $\hat{G}^\mathrm{SA}_{iiaa}$ corresponds to $\hat{G}_{i_\alpha i_\beta}^{a_\alpha a_\beta}$. 
Each adaptive VQE ansatz was constructed iteratively by selecting the operator with the largest energy gradient magnitude at each step until the convergence criterion was reached.

\subsection{Adaptive VQE calculations}

With these operator pools defined, we now compare their performance for representative molecular systems. 
Using the fermionic operator pool 
\(\{\hat{G}_{i_\sigma}^{a_\sigma},\hat{G}_{i_\sigma j_\tau}^{a_\sigma b_\tau},\hat{G}_{i_\sigma j_\tau}^{a_\tau b_\sigma}\}\) 
constrains the variational space to determinants that conserve particle number and spin projection. 
In contrast, the spin-adapted fermionic operator pool 
\(\{\hat{G}^\mathrm{SA}_{ia},\hat{G}^\mathrm{SA}_{ijab},\hat{G}^{\mathrm{SA}'}_{ijab}\}\) 
further enforces total spin conservation, effectively spanning the configuration-state function (CSF) space. 
Because the nonrelativistic, spin-free electronic Hamiltonian commutes with the total spin operator, it is expected that using spin-adapted generators should reduce the number of variational parameters required for convergence.

To quantify the effect of spin adaptation on circuit compactness, we compared the number of variational parameters required for convergence in adaptive VQE expansions for two systems: H$_2$O and BeH$_2$, both in the STO-3G basis set.

\begin{figure}[H]
    \centering
    \includegraphics[]{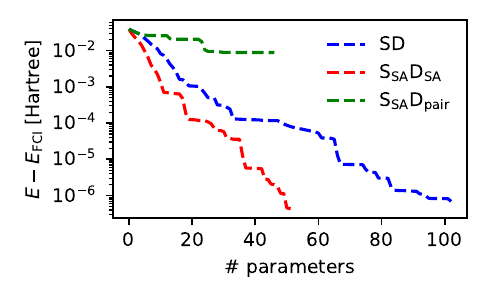}
    \caption{Adaptive VQE calculations for $\text{H}_2\text{O}$ with $\theta_\mathrm{HOH}=109^\circ$ and $R_\mathrm{OH}=0.977$ \AA. 
    SD denotes the fermionic operator pool, $\text{S}_\text{SA}\text{D}_\text{SA}$ the spin-adapted fermionic operator pool, and $\text{S}_\text{SA}\text{D}_\text{pair}$ the pool of spin-adapted singles and pair doubles.}
    \label{fig:h2o_adapt}
\end{figure}

Figure~\ref{fig:h2o_adapt} shows the convergence of adaptive VQE calculations for H$_2$O relative to the full configuration interaction (FCI) energy. 
Using the fermionic operator pool (blue dashed line) required 104 variational parameters to reach the gradient threshold of $10^{-5}$~a.u. 
In contrast, the spin-adapted fermionic operator pool (red dashed line) achieved convergence with only 54 parameters, demonstrating that enforcing total spin conservation leads to a more compact ansatz. 
Physically, this reduction arises because spin-adapted generators only couple configurations within a single spin manifold, eliminating redundant directions in parameter space.

The alternative pool of spin-adapted singles and pair doubles (green dashed line) also conserves total spin and consists of circuits that can be directly implemented with known gate structures. 
However, this ansatz frequently becomes trapped in local minima and, in this case, converged to a solution $8.8\times 10^{-3}$~Hartree above the FCI reference. 
A similar shortcoming of pair-doubles-based operator pools has been reported previously by Burton \textit{et~al.}\cite{Burton2023}

\begin{figure}[H]
    \centering
    \includegraphics[]{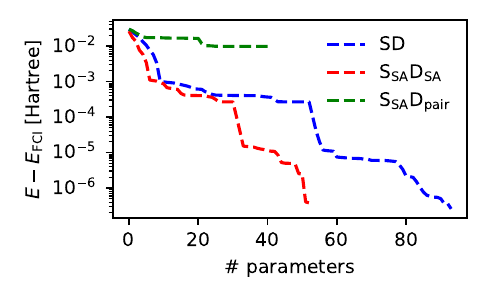}
    \caption{Adaptive VQE calculations for $\text{Be}\text{H}_2$ with $\theta_\mathrm{HBeH}=180^\circ$ and $R_\mathrm{BeH}=1.35$ \AA. 
    SD denotes the fermionic operator pool, $\text{S}_\text{SA}\text{D}_\text{SA}$ the spin-adapted fermionic operator pool, and $\text{S}_\text{SA}\text{D}_\text{pair}$ the pool of spin-adapted singles and pair doubles.}
    \label{fig:beh2_adapt}
\end{figure}

Figure~\ref{fig:beh2_adapt} presents analogous results for BeH$_2$. 
The same qualitative trend is observed: the fermionic operator pool required 95 parameters to converge, whereas the spin-adapted fermionic pool reached convergence with only 55 parameters. 
Again, the $\text{S}_\text{SA}\text{D}_\text{pair}$ operator pool became trapped in a local minimum and failed to reach the FCI energy. 
These results confirm that incorporating total spin symmetry systematically reduces the variational dimension required for accurate convergence.

\BC{We note that the reduction in the number of variational parameters does not automatically translate into a proportional reduction in circuit depth, because individual spin-adapted unitaries require more Pauli exponentials than non-symmetrized double excitations. Under the Jordan-Wigner mapping, a single non-symmetrized double excitation maps to $8$ Pauli strings, whereas the factorized spin-adapted operators require $48$ (seniority-2), $384$ (seniority-4 singlet), or $640$ (seniority-4 triplet) Pauli strings (Table~\ref{tab:num_pauli}). However, the roughly twofold reduction in the number of operators needed for convergence partially compensates for this increase. For H$_2$O, for example, the SD ansatz uses $104$ operators $\times$ $8$ Pauli strings $= 832$ total Pauli exponentials, while the $\text{S}_\text{SA}\text{D}_\text{SA}$ ansatz uses $54$ operators with a mixture of seniority types, giving a total of $8528$ Pauli exponentials.
Here, the number of Pauli exponentials is completely dominated by the 6 added $\hat{G}^\text{SA}_{ijab}$ (S) and 8 added $\hat{G}^{\text{SA}'}_{ijab}$ (T), accounting for $7424$ of the Pauli exponentials.
The net circuit depth depends on the distribution of operator types selected by the adaptive algorithm. In practice, the primary advantage of spin adaptation lies not in circuit-depth reduction, but in ensuring physical correctness of the wavefunction by eliminating spin contamination, and improving the convergence landscape.}

An important advantage of using spin-adapted operators is their ability to prevent variational collapse to states of different total spin. 
To illustrate this point, we investigated the O$_2$ molecule in its triplet configuration, where the risk of spin contamination is high.

The test system used the molecular orbitals obtained from a restricted open-shell Hartree--Fock (ROHF) triplet calculation. 
The reference state was chosen as the triplet CSF
\[
\frac{1}{\sqrt{2}}\big(\pi^*_x(\downarrow)\pi^*_y(\uparrow)+\pi^*_x(\uparrow)\pi^*_y(\downarrow)\big),
\]
corresponding to $m_s=0$.

\begin{figure}[H]
    \centering
    \includegraphics[]{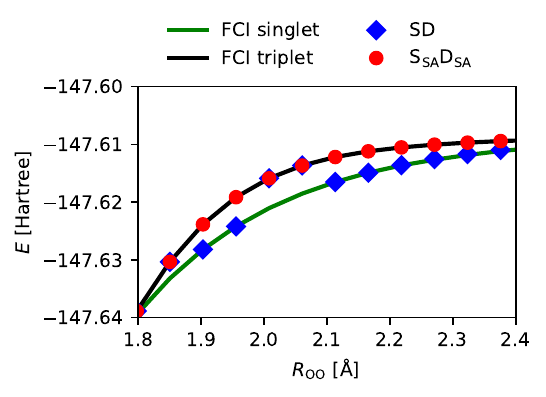}
    \caption{Adaptive VQE calculations for $\text{O}_2$ triplet. 
    SD denotes the fermionic operator pool, $\text{S}_\text{SA}\text{D}_\text{SA}$ the spin-adapted fermionic operator pool. 
    FCI triplet and singlet reference curves are shown for comparison.}
    \label{fig:o2_adapt}
\end{figure}

Figure~\ref{fig:o2_adapt} displays the potential energy curve of O$_2$ for bond lengths between 1.8 and 2.4~\AA. 
When using the spin-adapted fermionic operator pool ($\text{S}_\text{SA}\text{D}_\text{SA}$, red circles), the adaptive VQE energies follow the FCI triplet curve (black line) across all bond lengths. 
In contrast, using the conventional fermionic operator pool (SD, blue diamonds) leads to a collapse to the lower-energy singlet solution (green curve), even when starting from a triplet reference. 
This behavior arises because the SD operators do not conserve total spin and thus allow transitions between spin manifolds.

Previous studies have attempted to address this problem by adding penalty terms to the Hamiltonian during variational optimization\cite{Kuroiwa2021} or by applying spin-projection techniques\cite{Tsuchimochi2020}. 
However, these approaches increase computational cost either by complicating the energy functional or by requiring additional ancilla qubits. 
Using spin-adapted operators naturally preserves the desired spin symmetry without modifying the Hamiltonian or introducing extra qubits, offering a simple and physically consistent solution.

\section{Conclusions}

We have introduced a framework for factorizing spin-adapted unitary transformations into ordered products of exponentials within compact Lie algebras of symmetrized excitations. 
By employing the adjoint representation, the exponential of a sum of spin-adapted generators can be obtained via a low-dimensional nonlinear optimization over matrix exponentials, yielding numerically exact amplitudes without recourse to symbolic Baker--Campbell--Hausdorff expansions. 
This formulation preserves total spin and particle-number symmetries by construction and produces efficient, low-depth circuit realizations suitable for use in variational adaptive algorithms such as adaptive VQE schemes. 

The accompanying analysis of seniority-specific Lie algebras clarifies why such compact factorizations exist and how their algebraic structure influences parameter organization and numerical stability. 
In particular, the identification of small, closed algebras within each seniority sector accelerates convergence in the adjoint-space optimization and provides a foundation for constructing numerically stable, symmetry-consistent amplitude parametrizations. 
In benchmark adaptive VQE calculations, enforcing spin adaptation was shown to reduce the number of variational parameters by roughly a factor of two while maintaining exact spin symmetry throughout the optimization. 

Building on these results, the combined algebraic and numerical framework presented here establishes a general pathway for translating symmetry-adapted unitary transformations into implementable quantum circuits. 
Extending this approach to higher-seniority, spin--orbit-coupled, or alternative symmetry-adapted spaces may enable the design of new ansätze with greater expressivity at reduced circuit depth. 
More broadly, the adjoint-based formulation offers a principled route linking Lie-algebraic structure to variational algorithm design, providing a systematic strategy for constructing efficient, symmetry-preserving transformations for quantum simulation.
\BC{Because the factorization is purely algebraic, it may also find application in Trotterized Hamiltonian simulation, provided the Hamiltonian is partitioned into spin-conserving components whose Lie algebras admit a similar treatment.}

\section*{Acknowledgments}

 A.F.I. acknowledge financial support from the 
 Natural Sciences and Engineering Council of Canada (NSERC) and the M{\o}nsted Foundation for his research stay at the Department of Chemistry, Technical University of Denmark (DTU). 
 E.R.K. acknowledge the support of the Novo Nordisk Foundation (NNF) for the focused research project ``Hybrid Quantum Chemistry on Hybrid Quantum Computers'' (grant number  NNFSA220080996).

\bibliographystyle{unsrt}
\bibliography{literature}

@article{nonSurj,
  author  = {Dragomir {\v Z}, {\DJ}okovi{\'c} and Karl H. Hofmann},
  title   = {The Surjectivity Question for the Exponential Function of Real Lie Groups: A Status Report},
  journal = {Journal of Lie Theory},
  year    = {1997},
  volume  = {7},
  pages   = {171--199}
}

@article{ryabinkin2019constr,
  title={Constrained variational quantum eigensolver: Quantum computer search engine in the Fock space},
  author={Ryabinkin, Ilya G and Genin, Scott N and Izmaylov, Artur F},
  journal={J. Chem. Theory Comp.},
  volume={15},
  pages={249--255},
  year={2019}
}

@article{QCC,
  title={Qubit coupled-cluster method: A systematic approach to quantum chemistry on a quantum computer},
  author={Ryabinkin, Ilya G. and Yen, Tzu-Ching and Genin, Scott N. and Izmaylov, Artur F.},
  journal={J. Chem Theory Comput.},
  volume={14},
  pages={6317--6326},
  year={2018}
}

@article{ProjEA2019,
  title={Exact and approximate symmetry projectors for the electronic  structure problem on a quantum computer},
  author={Yen, Tzu-Ching and Lang, Robert A. and Izmaylov, Artur F.},
  journal={J. Chem. Phys.},
  volume={151},
  pages={164111},
  year={2019}
}

@article{iQCC,
  title={Iterative Qubit Coupled Cluster approach with efficient screening of generators},
  author={Ryabinkin, IG and Lang, RA and Genin, SN and Izmaylov, AF},
  journal={J. Chem. Theory Comput.},
  volume={16},
  number={2},
  pages={1055--1063},
  year={2020}
}

@article{UCC:rev,
  title={A Quantum Computing View on Unitary Coupled Cluster Theory},
  author={Anand, A and Schleich, P and Alperin-Lea, S and Jensen, PWK and Sim, S and D{\'\i}az-Tinoco, M and Kottmann, JS and Degroote, M and Izmaylov, AF and Aspuru-Guzik, A},
  journal={Chem. Soc. Rev.},
  volume={51},
  pages={1659--1684},
  year={2022}
}

@ARTICLE{Grimsley2019-zw,
  title     = "An adaptive variational algorithm for exact molecular
               simulations on a quantum computer",
  author    = "Grimsley, Harper R and Economou, Sophia E and Barnes, Edwin and
               Mayhall, Nicholas J",
  journal   = "Nat. Commun.",
  publisher = "Springer Science and Business Media LLC",
  volume    =  10,
  number    =  1,
  pages     = "3007",
  month     =  jul,
  year      =  2019,
  copyright = "https://creativecommons.org/licenses/by/4.0",
  language  = "en"
}

@ARTICLE{Hehre1969-ju,
  title     = "Self-consistent molecular-orbital methods. I. use of Gaussian
               expansions of Slater-type atomic orbitals",
  author    = "Hehre, W J and Stewart, R F and Pople, J A",
  journal   = "J. Chem. Phys.",
  publisher = "AIP Publishing",
  volume    =  51,
  number    =  6,
  pages     = "2657--2664",
  month     =  sep,
  year      =  1969,
  language  = "en"
}

@ARTICLE{Hehre1970-ym,
  title     = "Self-consistent molecular orbital methods. {IV}. Use of Gaussian
               expansions of Slater-type orbitals. Extension to second-row
               molecules",
  author    = "Hehre, W J and Ditchfield, R and Stewart, R F and Pople, J A",
  journal   = "J. Chem. Phys.",
  publisher = "AIP Publishing",
  volume    =  52,
  number    =  5,
  pages     = "2769--2773",
  month     =  mar,
  year      =  1970,
  language  = "en"
}

@misc{slowquant,
  title={Slow{Q}uant},
  author={Kjellgren, E. and Ziems, K. M.},
  year={2025},
  howpublished = {\url{https://github.com/erikkjellgren/SlowQuant}}
}

@misc{spinadapted,
  title={Spin{A}dapted},
  author={Paarth Jain},
  year={2025},
  howpublished = {\url{https://github.com/paarth7777/SpinAdapted}}
}

@ARTICLE{2020SciPy-NMeth,
  author  = {Virtanen, Pauli and Gommers, Ralf and Oliphant, Travis E. and
            Haberland, Matt and Reddy, Tyler and Cournapeau, David and
            Burovski, Evgeni and Peterson, Pearu and Weckesser, Warren and
            Bright, Jonathan and {van der Walt}, St{\'e}fan J. and
            Brett, Matthew and Wilson, Joshua and Millman, K. Jarrod and
            Mayorov, Nikolay and Nelson, Andrew R. J. and Jones, Eric and
            Kern, Robert and Larson, Eric and Carey, C J and
            Polat, {\.I}lhan and Feng, Yu and Moore, Eric W. and
            {VanderPlas}, Jake and Laxalde, Denis and Perktold, Josef and
            Cimrman, Robert and Henriksen, Ian and Quintero, E. A. and
            Harris, Charles R. and Archibald, Anne M. and
            Ribeiro, Ant{\^o}nio H. and Pedregosa, Fabian and
            {van Mulbregt}, Paul and {SciPy 1.0 Contributors}},
  title   = {{{SciPy} 1.0: Fundamental Algorithms for Scientific
            Computing in Python}},
  journal = {Nature Methods},
  year    = {2020},
  volume  = {17},
  pages   = {261--272},
  adsurl  = {https://rdcu.be/b08Wh},
  doi     = {10.1038/s41592-019-0686-2},
}

@article{Anselmetti_2021,
doi = {10.1088/1367-2630/ac2cb3},
url = {https://doi.org/10.1088/1367-2630/ac2cb3},
year = {2021},
month = {nov},
publisher = {IOP Publishing},
volume = {23},
number = {11},
pages = {113010},
author = {Anselmetti, Gian-Luca R and Wierichs, David and Gogolin, Christian and Parrish, Robert M},
title = {Local, expressive, quantum-number-preserving VQE ansätze for fermionic systems},
journal = {New Journal of Physics}
}

@article{Qubit-ADAPT,
   title={Qubit-ADAPT-VQE: An Adaptive Algorithm for Constructing Hardware-Efficient Ansätze on a Quantum Processor},
   volume={2},
   ISSN={2691-3399},
   url={http://dx.doi.org/10.1103/PRXQuantum.2.020310},
   DOI={10.1103/prxquantum.2.020310},
   number={2},
   journal={PRX Quantum},
   publisher={American Physical Society (APS)},
   author={Tang, Ho Lun and Shkolnikov, V.O. and Barron, George S. and Grimsley, Harper R. and Mayhall, Nicholas J. and Barnes, Edwin and Economou, Sophia E.},
   year={2021},
   month=apr }

@article{Burton2023,
  title = {Exact electronic states with shallow quantum circuits from global optimisation},
  volume = {9},
  ISSN = {2056-6387},
  url = {http://dx.doi.org/10.1038/s41534-023-00744-2},
  DOI = {10.1038/s41534-023-00744-2},
  number = {1},
  journal = {npj Quantum Information},
  publisher = {Springer Science and Business Media LLC},
  author = {Burton,  Hugh G. A. and Marti-Dafcik,  Daniel and Tew,  David P. and Wales,  David J.},
  year = {2023},
  month = jul 
}

@article{Kuroiwa2021,
  title = {Penalty methods for a variational quantum eigensolver},
  volume = {3},
  ISSN = {2643-1564},
  url = {http://dx.doi.org/10.1103/PhysRevResearch.3.013197},
  DOI = {10.1103/physrevresearch.3.013197},
  number = {1},
  journal = {Physical Review Research},
  publisher = {American Physical Society (APS)},
  author = {Kuroiwa,  Kohdai and Nakagawa,  Yuya O.},
  year = {2021},
  month = feb 
}

@article{Tsuchimochi2020,
  title = {Spin-projection for quantum computation: A low-depth approach to strong correlation},
  volume = {2},
  ISSN = {2643-1564},
  url = {http://dx.doi.org/10.1103/PhysRevResearch.2.043142},
  DOI = {10.1103/physrevresearch.2.043142},
  number = {4},
  journal = {Physical Review Research},
  publisher = {American Physical Society (APS)},
  author = {Tsuchimochi,  Takashi and Mori,  Yuto and Ten-no,  Seiichiro L.},
  year = {2020},
  month = oct 
}

@article{Bertels2022,
  author = {Bertels, Luke W. and Grimsley, Harper R. and Economou, Sophia E. and Barnes, Edwin and Mayhall, Nicholas J.},
  title = {Symmetry Breaking Slows Convergence of the ADAPT Variational Quantum Eigensolver},
  journal = {J. Chem. Theory Comput.},
  volume = {18},
  number = {11},
  pages = {6656--6669},
  year = {2022},
  doi = {10.1021/acs.jctc.2c00709}
}

@ARTICLE{Magoulas2025-bz,
  title         = "Spin-adapted fermionic unitaries: From Lie algebras to
                   compact quantum circuits",
  author        = "Magoulas, Ilias and Evangelista, Francesco A",
  month         =  nov,
  year          =  2025,
  copyright     = "http://arxiv.org/licenses/nonexclusive-distrib/1.0/",
  archivePrefix = "arXiv",
  primaryClass  = "quant-ph",
  eprint        = "2511.13485"
}

\clearpage
\appendix
\onecolumngrid

\section{Seniority 4, intermediate singlet}
\label{app:sen4s}

This algebra is generated by 
$A_1=\hat G^{a_\alpha b_\beta}_{i_\alpha j_\beta}$, $A_2=\hat G^{a_\beta b_\alpha}_{i_\beta j_\alpha}$, 
$A_3=\hat G^{a_\alpha b_\beta}_{i_\beta j_\alpha}$, $A_4=\hat G^{a_\beta b_\alpha}_{i_\alpha j_\beta}$ from 
the symmetrized generator of \eq{eq:4s}. 
The generated Lie algebra has dimension $28$, $\mathcal A_{S4S}= \{A_p\}_{p=1}^{28}$, with the following 
generators:

{\it First-Level Commutators:}
\begin{align*}
    A_5 &= \left(1 - \hat{n}(a_\alpha) - \hat{n}(b_\beta)\right)\hat{G}_{i_\alpha j_\beta}^{i_\beta j_\alpha} \\
    A_6 &= \left(1 - \hat{n}(i_\alpha) - \hat{n}(j_\beta)\right)\hat{G}_{a_\alpha b_\beta}^{a_\beta b_\alpha} \\
    A_7 &= -\left(1 - \hat{n}(i_\beta) - \hat{n}(j_\alpha)\right)\hat{G}_{a_\alpha b_\beta}^{a_\beta b_\alpha} \\
    A_8 &= -\left(1 - \hat{n}(a_\beta) - \hat{n}(b_\alpha)\right)\hat{G}_{i_\alpha j_\beta}^{i_\beta j_\alpha}
\end{align*}

{\it Second-Level Commutators:}
\begin{align*}
    A_9 &= -\left(1 - \hat{n}(i_\alpha) - \hat{n}(j_\beta) + 2\hat{n}(i_\alpha)\hat{n}(j_\beta)\right)\hat{G}_{i_\beta j_\alpha}^{a_\alpha b_\beta} \\
    A_{10} &= -\left(1 - \hat{n}(a_\alpha) - \hat{n}(b_\beta) + 2\hat{n}(a_\alpha)\hat{n}(b_\beta)\right)\hat{G}_{i_\alpha j_\beta}^{a_\beta b_\alpha} \\
    A_{11} &= \left(1-\hat{n}(i_\beta) - \hat{n}(j_\alpha) - \hat{n}(a_\alpha) - \hat{n}(b_\beta) + \hat{n}(i_\beta)\hat{n}(b_\beta) \right. \\
    &\quad \left. + \hat{n}(i_\beta)\hat{n}(a_\alpha) + \hat{n}(j_\alpha)\hat{n}(b_\beta) + \hat{n}(j_\alpha)\hat{n}(a_\alpha)\right)\hat{G}_{i_\alpha j_\beta}^{a_\beta b_\alpha} \\
    A_{12} &= \left(1- \hat{n}(i_\alpha) - \hat{n}(j_\beta) - \hat{n}(a_\beta) - \hat{n}(b_\alpha) + \hat{n}(i_\alpha)\hat{n}(a_\beta) \right. \\
    &\quad \left. + \hat{n}(i_\alpha)\hat{n}(b_\alpha) + \hat{n}(j_\beta)\hat{n}(a_\beta) + \hat{n}(j_\beta)\hat{n}(b_\alpha)\right)\hat{G}_{i_\beta j_\alpha}^{a_\alpha b_\beta} \\
    A_{13} &= -\left(1- \hat{n}(a_\beta) - \hat{n}(b_\alpha) + 2\hat{n}(a_\beta)\hat{n}(b_\alpha)\right)\hat{G}_{i_\beta j_\alpha}^{a_\alpha b_\beta} \\
    A_{14} &= -\left(1- \hat{n}(i_\beta) - \hat{n}(j_\alpha) + 2\hat{n}(i_\beta)\hat{n}(j_\alpha)\right)\hat{G}_{i_\alpha j_\beta}^{a_\beta b_\alpha} \\
    A_{15} &= \left(1- \hat{n}(i_\beta) - \hat{n}(j_\alpha) + 2\hat{n}(i_\beta)\hat{n}(j_\alpha)\right)\hat{G}_{i_\alpha j_\beta}^{a_\alpha b_\beta} \\
    A_{16} &= -\left(1 - \hat{n}(i_\alpha) - \hat{n}(j_\beta) - \hat{n}(a_\alpha) - \hat{n}(b_\beta) + \hat{n}(i_\alpha)\hat{n}(a_\alpha) \right. \\
    &\quad \left. + \hat{n}(i_\alpha)\hat{n}(b_\beta) + \hat{n}(j_\beta)\hat{n}(a_\alpha) + \hat{n}(j_\beta)\hat{n}(b_\beta)\right)\hat{G}_{i_\beta j_\alpha}^{a_\beta b_\alpha} \\
    A_{17} &= \left(1 - \hat{n}(a_\alpha) - \hat{n}(b_\beta) + 2\hat{n}(a_\alpha)\hat{n}(b_\beta) \right)\hat{G}_{i_\beta j_\alpha}^{a_\beta b_\alpha} \\
    A_{18} &= -\left(1 - \hat{n}(i_\beta) - \hat{n}(j_\alpha) - \hat{n}(a_\beta) - \hat{n}(b_\alpha) + \hat{n}(i_\beta)\hat{n}(a_\beta) \right. \\
    &\quad \left. + \hat{n}(i_\beta)\hat{n}(b_\alpha) + \hat{n}(j_\alpha)\hat{n}(a_\beta) + \hat{n}(j_\alpha)\hat{n}(b_\alpha)\right)\hat{G}_{i_\alpha j_\beta}^{a_\alpha b_\beta} \\
    A_{19} &= \left(1 - \hat{n}(a_\beta) - \hat{n}(b_\alpha) + 2\hat{n}(a_\beta)\hat{n}(b_\alpha)\right)\hat{G}_{i_\alpha j_\beta}^{a_\alpha b_\beta} \\
    A_{20} &= \left(1 - \hat{n}(i_\alpha) - \hat{n}(j_\beta) + 2\hat{n}(i_\alpha)\hat{n}(j_\beta)\right)\hat{G}_{i_\beta j_\alpha}^{a_\beta b_\alpha}
\end{align*}

{\it Third-Level Commutators:}
\begin{align*}
    A_{21} &= \left(1 - \hat{n}(i_\alpha) - \hat{n}(i_\beta) - \hat{n}(j_\alpha) - \hat{n}(j_\beta) + \hat{n}(i_\alpha)\hat{n}(i_\beta) + \hat{n}(i_\alpha)\hat{n}(j_\alpha) \right. \\
    &\quad \left. + \hat{n}(i_\beta)\hat{n}(j_\beta) + \hat{n}(j_\alpha)\hat{n}(j_\beta) + 2\hat{n}(i_\alpha)\hat{n}(j_\beta) - 2\hat{n}(i_\alpha)\hat{n}(i_\beta)\hat{n}(j_\beta) \right. \\
    &\quad \left. - 2\hat{n}(i_\alpha)\hat{n}(j_\alpha)\hat{n}(j_\beta)\right)\hat{G}_{a_\alpha b_\beta}^{a_\beta b_\alpha} \\
    A_{22} &= \left(1 - \hat{n}(a_\alpha) - \hat{n}(a_\beta) - \hat{n}(b_\alpha) - \hat{n}(b_\beta) + \hat{n}(a_\alpha)\hat{n}(a_\beta) + \hat{n}(a_\alpha)\hat{n}(b_\alpha) \right. \\
    &\quad \left. + \hat{n}(a_\beta)\hat{n}(b_\beta) + \hat{n}(b_\alpha)\hat{n}(b_\beta) + 2\hat{n}(a_\alpha)\hat{n}(b_\beta) - 2\hat{n}(a_\alpha)\hat{n}(a_\beta)\hat{n}(b_\beta) \right. \\
    &\quad \left. - 2\hat{n}(a_\alpha)\hat{n}(b_\alpha)\hat{n}(b_\beta)\right)\hat{G}_{i_\alpha j_\beta}^{i_\beta j_\alpha} \\
    A_{23} &= -\left(1 - \hat{n}(a_\alpha) - \hat{n}(a_\beta) - \hat{n}(b_\alpha) - \hat{n}(b_\beta) + \hat{n}(a_\alpha)\hat{n}(a_\beta) + \hat{n}(a_\alpha)\hat{n}(b_\alpha) \right. \\
    &\quad \left. + \hat{n}(a_\beta)\hat{n}(b_\beta) + \hat{n}(b_\alpha)\hat{n}(b_\beta) + 2\hat{n}(a_\beta)\hat{n}(b_\alpha) - 2\hat{n}(a_\alpha)\hat{n}(a_\beta)\hat{n}(b_\alpha) \right. \\
    &\quad \left. - 2\hat{n}(a_\beta)\hat{n}(b_\alpha)\hat{n}(b_\beta)\right)\hat{G}_{i_\alpha j_\beta}^{i_\beta j_\alpha} \\
    A_{24} &= -\left(1 - \hat{n}(i_\alpha) - \hat{n}(i_\beta) - \hat{n}(j_\alpha) - \hat{n}(j_\beta) + \hat{n}(i_\alpha)\hat{n}(i_\beta) + \hat{n}(i_\alpha)\hat{n}(j_\alpha) \right. \\
    &\quad \left. + \hat{n}(i_\beta)\hat{n}(j_\beta) + \hat{n}(j_\alpha)\hat{n}(j_\beta) + 2\hat{n}(i_\beta)\hat{n}(j_\alpha) - 2\hat{n}(i_\alpha)\hat{n}(i_\beta)\hat{n}(j_\alpha) \right. \\
    &\quad \left. - 2\hat{n}(i_\beta)\hat{n}(j_\alpha)\hat{n}(j_\beta)\right)\hat{G}_{a_\alpha b_\beta}^{a_\beta b_\alpha}
\end{align*}

{\it Fourth-Level Commutators:}
\begin{align*}
    A_{25} &= \left(1 - \hat{n}(i_\alpha) - \hat{n}(j_\beta) - \hat{n}(a_\beta) - \hat{n}(b_\alpha) + \hat{n}(i_\alpha)\hat{n}(b_\alpha) + \hat{n}(i_\alpha)\hat{n}(a_\beta) \right. \\
    &\quad \left. + \hat{n}(j_\beta)\hat{n}(b_\alpha) + \hat{n}(j_\beta)\hat{n}(a_\beta) + 2\hat{n}(i_\alpha)\hat{n}(j_\beta) + 2\hat{n}(a_\beta)\hat{n}(b_\alpha) \right. \\
    &\quad \left. - 2\hat{n}(i_\alpha)\hat{n}(a_\beta)\hat{n}(b_\alpha) - 2\hat{n}(j_\beta)\hat{n}(a_\beta)\hat{n}(b_\alpha) - 2\hat{n}(i_\alpha)\hat{n}(j_\beta)\hat{n}(b_\alpha) \right. \\
    &\quad \left. - 2\hat{n}(i_\alpha)\hat{n}(j_\beta)\hat{n}(a_\beta) + 4\hat{n}(i_\alpha)\hat{n}(j_\beta)\hat{n}(a_\beta)\hat{n}(b_\alpha)\right)\hat{G}_{i_\beta j_\alpha}^{a_\alpha b_\beta} \\
    A_{26} &= \left(1 - \hat{n}(i_\beta) - \hat{n}(j_\alpha) - \hat{n}(a_\alpha) - \hat{n}(b_\beta) + \hat{n}(j_\alpha)\hat{n}(a_\alpha) + \hat{n}(j_\alpha)\hat{n}(b_\beta) \right. \\
    &\quad \left. + \hat{n}(i_\beta)\hat{n}(a_\alpha) + \hat{n}(i_\beta)\hat{n}(b_\beta) + 2\hat{n}(i_\beta)\hat{n}(j_\alpha) + 2\hat{n}(a_\alpha)\hat{n}(b_\beta) \right. \\
    &\quad \left. - 2\hat{n}(i_\beta)\hat{n}(j_\alpha)\hat{n}(a_\alpha) - 2\hat{n}(i_\beta)\hat{n}(j_\alpha)\hat{n}(b_\beta) - 2\hat{n}(j_\alpha)\hat{n}(a_\alpha)\hat{n}(b_\beta) \right. \\
    &\quad \left. - 2\hat{n}(i_\beta)\hat{n}(a_\alpha)\hat{n}(b_\beta) + 4\hat{n}(i_\beta)\hat{n}(j_\alpha)\hat{n}(a_\alpha)\hat{n}(b_\beta)\right)\hat{G}_{i_\alpha j_\beta}^{a_\beta b_\alpha}
\end{align*}
\begin{align*}    
    A_{27} &= -\left(1 - \hat{n}(i_\alpha) - \hat{n}(j_\beta) - \hat{n}(a_\alpha) - \hat{n}(b_\beta) + \hat{n}(i_\alpha)\hat{n}(a_\alpha) + \hat{n}(i_\alpha)\hat{n}(b_\beta) \right. \\
    &\quad \left. + \hat{n}(j_\beta)\hat{n}(a_\alpha) + \hat{n}(j_\beta)\hat{n}(b_\beta) + 2\hat{n}(i_\alpha)\hat{n}(j_\beta) + 2\hat{n}(a_\alpha)\hat{n}(b_\beta) \right. \\
    &\quad \left. - 2\hat{n}(i_\alpha)\hat{n}(a_\alpha)\hat{n}(b_\beta) - 2\hat{n}(j_\beta)\hat{n}(a_\alpha)\hat{n}(b_\beta) - 2\hat{n}(i_\alpha)\hat{n}(j_\beta)\hat{n}(a_\alpha) \right. \\
    &\quad \left. - 2\hat{n}(i_\alpha)\hat{n}(j_\beta)\hat{n}(b_\beta) + 4\hat{n}(i_\alpha)\hat{n}(j_\beta)\hat{n}(a_\alpha)\hat{n}(b_\beta)\right)\hat{G}_{i_\beta j_\alpha}^{a_\beta b_\alpha} \\
    A_{28} &= -\left(1 - \hat{n}(i_\beta) - \hat{n}(j_\alpha) - \hat{n}(a_\beta) - \hat{n}(b_\alpha) + \hat{n}(i_\beta)\hat{n}(a_\beta) + \hat{n}(i_\beta)\hat{n}(b_\alpha) \right. \\
    &\quad \left. + \hat{n}(j_\alpha)\hat{n}(a_\beta) + \hat{n}(j_\alpha)\hat{n}(b_\alpha) + 2\hat{n}(i_\beta)\hat{n}(j_\alpha) + 2\hat{n}(a_\beta)\hat{n}(b_\alpha) \right. \\
    &\quad \left. - 2\hat{n}(i_\beta)\hat{n}(a_\beta)\hat{n}(b_\alpha) - 2\hat{n}(j_\alpha)\hat{n}(a_\beta)\hat{n}(b_\alpha) - 2\hat{n}(i_\beta)\hat{n}(j_\alpha)\hat{n}(a_\beta) \right. \\
    &\quad \left. - 2\hat{n}(i_\beta)\hat{n}(j_\alpha)\hat{n}(b_\alpha) + 4\hat{n}(i_\beta)\hat{n}(j_\alpha)\hat{n}(a_\beta)\hat{n}(b_\alpha)\right)\hat{G}_{i_\alpha j_\beta}^{a_\alpha b_\beta}
\end{align*}

The commutation relations of these are given in Table~\ref{tab:commutation}.

\subsection*{Commuting groups}

The groups of commuting generators are,

\begin{align}
s&=\{1,2,15,16,17,18,19,20,27,28\}\\
s&=\{3,4,9,10,11,12,13,14,25,26\}\\
s&=\{5,6,7,8,21,22,23,24\}
\end{align}

\begin{sidewaystable}
    \centering
    \caption{Commutation Table for the 28-Element Seniority-4 Lie Algebra $[A_i, A_j]$}
    \label{tab:commutation}
    % \tiny % Use a very small font size for readability
    \setlength{\tabcolsep}{3.5pt} % Adjust column spacing
    \scalebox{0.8}{
    \begin{tabular}{|r|*{28}{c|}}
        \hline
        \textbf{[i, j]} & \textbf{1} & \textbf{2} & \textbf{3} & \textbf{4} & \textbf{5} & \textbf{6} & \textbf{7} & \textbf{8} & \textbf{9} & \textbf{10} & \textbf{11} & \textbf{12} & \textbf{13} & \textbf{14} & \textbf{15} & \textbf{16} & \textbf{17} & \textbf{18} & \textbf{19} & \textbf{20} & \textbf{21} & \textbf{22} & \textbf{23} & \textbf{24} & \textbf{25} & \textbf{26} & \textbf{27} & \textbf{28} \\
        \hline
\textbf{1} & 0 & 0 & 5 & 6 & 9 & 10 & 11 & 12 & -5 & -6 & 21 & 22 & 23 & 24 & 0 & 0 & 0 & 0 & 0 & 0 & -11 & -12 & 25 & 26 & -23 & -24 & 0 & 0 \\ \hline
\textbf{2} & 0 & 0 & 7 & 8 & 11 & 12 & 13 & 14 & 21 & 22 & 23 & 24 & -7 & -8 & 0 & 0 & 0 & 0 & 0 & 0 & 25 & 26 & -11 & -12 & -21 & -22 & 0 & 0 \\ \hline
\textbf{3} & -5 & -7 & 0 & 0 & 15 & 16 & 17 & 18 & 0 & 0 & 0 & 0 & 0 & 0 & -5 & 24 & -7 & 22 & 23 & 21 & 27 & -18 & 28 & -16 & 0 & 0 & -21 & -23 \\ \hline
\textbf{4} & -6 & -8 & 0 & 0 & 16 & 19 & 18 & 20 & 0 & 0 & 0 & 0 & 0 & 0 & 24 & 23 & 22 & 21 & -6 & -8 & -18 & 27 & -16 & 28 & 0 & 0 & -22 & -24 \\ \hline
\textbf{5} & -9 & -11 & -15 & -16 & 0 & 0 & 0 & 0 & 15 & 16 & -27 & 18 & -28 & 16 & -9 & 26 & -11 & -12 & 25 & -11 & 0 & 0 & 0 & 0 & 28 & -16 & 11 & -25 \\ \hline
\textbf{6} & -10 & -12 & -16 & -19 & 0 & 0 & 0 & 0 & 16 & 19 & 18 & -27 & 16 & -28 & 26 & 25 & -12 & -11 & -10 & -12 & 0 & 0 & 0 & 0 & -16 & 28 & 12 & -26 \\ \hline
\textbf{7} & -11 & -13 & -17 & -18 & 0 & 0 & 0 & 0 & -27 & 18 & -28 & 16 & 17 & 18 & -11 & -12 & -13 & 26 & -11 & 25 & 0 & 0 & 0 & 0 & 27 & -18 & -25 & 11 \\ \hline
\textbf{8} & -12 & -14 & -18 & -20 & 0 & 0 & 0 & 0 & 18 & -27 & 16 & -28 & 18 & 20 & -12 & -11 & 26 & 25 & -12 & -14 & 0 & 0 & 0 & 0 & -18 & 27 & -26 & 12 \\ \hline
\textbf{9} & 5 & -21 & 0 & 0 & -15 & -16 & 27 & -18 & 0 & 0 & 0 & 0 & 0 & 0 & 5 & -24 & -21 & -22 & -23 & -21 & -27 & 18 & -28 & 16 & 0 & 0 & 21 & 23 \\ \hline
\textbf{10} & 6 & -22 & 0 & 0 & -16 & -19 & -18 & 27 & 0 & 0 & 0 & 0 & 0 & 0 & -24 & -23 & -22 & -21 & 6 & -22 & 18 & -27 & 16 & -28 & 0 & 0 & 22 & 24 \\ \hline
\textbf{11} & -21 & -23 & 0 & 0 & 27 & -18 & 28 & -16 & 0 & 0 & 0 & 0 & 0 & 0 & -21 & -22 & -23 & -24 & -21 & -23 & -28 & 16 & -27 & 18 & 0 & 0 & 23 & 21 \\ \hline
\textbf{12} & -22 & -24 & 0 & 0 & -18 & 27 & -16 & 28 & 0 & 0 & 0 & 0 & 0 & 0 & -22 & -21 & -24 & -23 & -22 & -24 & 16 & -28 & 18 & -27 & 0 & 0 & 24 & 22 \\ \hline
\textbf{13} & -23 & 7 & 0 & 0 & 28 & -16 & -17 & -18 & 0 & 0 & 0 & 0 & 0 & 0 & -23 & -24 & 7 & -22 & -23 & -21 & -27 & 18 & -28 & 16 & 0 & 0 & 21 & 23 \\ \hline
\textbf{14} & -24 & 8 & 0 & 0 & -16 & 28 & -18 & -20 & 0 & 0 & 0 & 0 & 0 & 0 & -24 & -23 & -22 & -21 & -24 & 8 & 18 & -27 & 16 & -28 & 0 & 0 & 22 & 24 \\ \hline
\textbf{15} & 0 & 0 & 5 & -24 & 9 & -26 & 11 & 12 & -5 & 24 & 21 & 22 & 23 & 24 & 0 & 0 & 0 & 0 & 0 & 0 & -11 & -12 & 25 & 26 & -23 & -24 & 0 & 0 \\ \hline
\textbf{16} & 0 & 0 & -24 & -23 & -26 & -25 & 12 & 11 & 24 & 23 & 22 & 21 & 24 & 23 & 0 & 0 & 0 & 0 & 0 & 0 & -12 & -11 & 26 & 25 & -24 & -23 & 0 & 0 \\ \hline
\textbf{17} & 0 & 0 & 7 & -22 & 11 & 12 & 13 & -26 & 21 & 22 & 23 & 24 & -7 & 22 & 0 & 0 & 0 & 0 & 0 & 0 & 25 & 26 & -11 & -12 & -21 & -22 & 0 & 0 \\ \hline
\textbf{18} & 0 & 0 & -22 & -21 & 12 & 11 & -26 & -25 & 22 & 21 & 24 & 23 & 22 & 21 & 0 & 0 & 0 & 0 & 0 & 0 & 26 & 25 & -12 & -11 & -22 & -21 & 0 & 0 \\ \hline
\textbf{19} & 0 & 0 & -23 & 6 & -25 & 10 & 11 & 12 & 23 & -6 & 21 & 22 & 23 & 24 & 0 & 0 & 0 & 0 & 0 & 0 & -11 & -12 & 25 & 26 & -23 & -24 & 0 & 0 \\ \hline
\textbf{20} & 0 & 0 & -21 & 8 & 11 & 12 & -25 & 14 & 21 & 22 & 23 & 24 & 21 & -8 & 0 & 0 & 0 & 0 & 0 & 0 & 25 & 26 & -11 & -12 & -21 & -22 & 0 & 0 \\ \hline
\textbf{21} & 11 & -25 & -27 & 18 & 0 & 0 & 0 & 0 & 27 & -18 & 28 & -16 & 27 & -18 & 11 & 12 & -25 & -26 & 11 & -25 & 0 & 0 & 0 & 0 & -27 & 18 & 25 & -11 \\ \hline
\textbf{22} & 12 & -26 & 18 & -27 & 0 & 0 & 0 & 0 & -18 & 27 & -16 & 28 & -18 & 27 & 12 & 11 & -26 & -25 & 12 & -26 & 0 & 0 & 0 & 0 & 18 & -27 & 26 & -12 \\ \hline
\textbf{23} & -25 & 11 & -28 & 16 & 0 & 0 & 0 & 0 & 28 & -16 & 27 & -18 & 28 & -16 & -25 & -26 & 11 & 12 & -25 & 11 & 0 & 0 & 0 & 0 & -28 & 16 & -11 & 25 \\ \hline
\textbf{24} & -26 & 12 & 16 & -28 & 0 & 0 & 0 & 0 & -16 & 28 & -18 & 27 & -16 & 28 & -26 & -25 & 12 & 11 & -26 & 12 & 0 & 0 & 0 & 0 & 16 & -28 & -12 & 26 \\ \hline
\textbf{25} & 23 & 21 & 0 & 0 & -28 & 16 & -27 & 18 & 0 & 0 & 0 & 0 & 0 & 0 & 23 & 24 & 21 & 22 & 23 & 21 & 27 & -18 & 28 & -16 & 0 & 0 & -21 & -23 \\ \hline
\textbf{26} & 24 & 22 & 0 & 0 & 16 & -28 & 18 & -27 & 0 & 0 & 0 & 0 & 0 & 0 & 24 & 23 & 22 & 21 & 24 & 22 & -18 & 27 & -16 & 28 & 0 & 0 & -22 & -24 \\ \hline
\textbf{27} & 0 & 0 & 21 & 22 & -11 & -12 & 25 & 26 & -21 & -22 & -23 & -24 & -21 & -22 & 0 & 0 & 0 & 0 & 0 & 0 & -25 & -26 & 11 & 12 & 21 & 22 & 0 & 0 \\ \hline
\textbf{28} & 0 & 0 & 23 & 24 & 25 & 26 & -11 & -12 & -23 & -24 & -21 & -22 & -23 & -24 & 0 & 0 & 0 & 0 & 0 & 0 & 11 & 12 & -25 & -26 & 23 & 24 & 0 & 0 \\ \hline
    \end{tabular}
    }
\end{sidewaystable}

%Structural analysis shows that the semi-simple part of $\mathcal A_{S4S}$ contains 4 ideals. 
\subsection*{Algebraic structure}

Let us denote $\mathfrak s=[\mathcal A_{S4S},\mathcal A_{S4S}]$  
the derived algebra. The restriction of the Killing form to $\mathfrak s$ is negative-definite and nondegenerate, i.e.\ $\mathfrak s$ is of compact type. Additionally, 
$\mathfrak s$ has no nontrivial proper ideals.
From the commutator table, $\mathfrak s =\mathrm{span}\{A_5,A_6,\dots,A_{28}\}$, $\dim \mathfrak s =24$.
Consequently, $\mathcal A_{S4S}$ admits a Levi decomposition
$\mathcal A_{S4S}=\mathfrak s\oplus \mathfrak r$, 
$\mathfrak r=Z(\mathcal A_{S4S})$ solvable radical, $\dim=4$.

While no single $A_i$ is central, the center $Z(\mathcal A_{S4S})$ 
is $4$-dimensional, with the following convenient basis of linear combinations:
\begin{align*}
Z_1&=A_3+A_9+A_{13}+A_{25},\\
Z_2&=A_4+A_{10}+A_{14}+A_{26},\\
Z_3&=-\,A_2+A_{17}+A_{20}+A_{27},\\
Z_4&=-\,A_1+A_{15}+A_{19}+A_{28},
\end{align*}
thus, $Z(\mathcal A_{S4S})=\mathrm{span}\{Z_1,Z_2,Z_3,Z_4\}\cong \mathbb R^4$.
The central directions have the following explicit forms:
\begin{align*}
Z_{1} &= \big(\hat{n}(i_\alpha) - \hat{n}(j_\beta)\big)^{2}
        \big(\hat{n}(a_\beta) - \hat{n}(b_\alpha)\big)^{2}\,
        \hat{G}_{i_\beta j_\alpha}^{ a_\alpha b_\beta} \\[4pt]
Z_{2} &= \big(\hat{n}(j_\alpha) - \hat{n}(i_\beta)\big)^{2}
        \big(\hat{n}(a_\alpha) - \hat{n}(b_\beta)\big)^{2}\,
        \hat{G}_{i_\alpha j_\beta}^{ a_\beta b_\alpha} \\[4pt]
Z_{3} &= -\Big(
          \hat{n}(i_\alpha)\hat{n}(a_\alpha)\big(1-\hat{n}(b_\beta)\big)\big(1-\hat{n}(j_\beta)\big)
        + \hat{n}(j_\beta)\hat{n}(b_\beta)\big(1-\hat{n}(i_\alpha)\big)\big(1-\hat{n}(a_\alpha)\big)
        \Big)\,
        \hat{G}_{i_\beta j_\alpha}^{ a_\beta b_\alpha} \\[4pt]
Z_{4} &= -\Big(
          \hat{n}(i_\beta)\hat{n}(a_\beta)\big(1-\hat{n}(b_\alpha)\big)\big(1-\hat{n}(j_\alpha)\big)
        + \hat{n}(j_\alpha)\hat{n}(b_\alpha)\big(1-\hat{n}(i_\beta)\big)\big(1-\hat{n}(a_\beta)\big)
        \Big)\,
        \hat{G}_{i_\alpha j_\beta}^{ a_\alpha b_\beta}
\end{align*}

For $i=1,2,3,4$ let $\mathcal A_i$ denote the $4$-tuple consisting of $A_i$ together with its three ``partners'':
\begin{align*}
\mathcal A_1&=\{A_1,A_{15},A_{19},A_{28}\},&
\mathcal A_2&=\{A_2,A_{17},A_{20},A_{27}\},\\
\mathcal A_3&=\{A_3,A_{9},A_{13},A_{25}\},&
\mathcal A_4&=\{A_4,A_{10},A_{14},A_{26}\}.
\end{align*}
Within each $\mathcal A_i$ all basis elements commute pairwise, so $\langle\mathcal A_i\rangle$ is a $4$-dimensional abelian subalgebra. 
However, only the specific linear combination displayed above lies in the center; the remaining vectors in each $\mathcal A_i$ are not central.

\section{Seniority 4, intermediate triplet}
\label{app:sen4t}

This 84-dimensional Lie algebra originates from the commutator closure of 6 elementary generators (the closure produces 84 linearly independent as fermionic operators) 
\begin{align*}
A_{1} &=  \hat{G}_{i_\alpha j_\beta}^{ a_\alpha b_\beta}\\
A_{2} &=  \hat{G}_{i_\beta j_\alpha}^{ a_\beta b_\alpha}\\
A_{3} &=  \hat{G}_{i_\beta j_\alpha}^{ a_\alpha b_\beta}\\
A_{4} &=  \hat{G}_{i_\alpha j_\beta}^{ a_\beta b_\alpha}\\
A_{5} &=  \hat{G}_{i_\alpha j_\alpha}^{ a_\alpha b_\alpha}\\
A_{6} &=  \hat{G}_{i_\beta j_\beta}^{ a_\beta b_\beta}
\end{align*}
In what follows we list all other generators of this algebra. 

{\it First-Level Commutators:}
\begin{align*}
A_{7} &=  (1 - \hat{n}(a_\alpha) - \hat{n}(b_\beta))\hat{G}_{i_\alpha j_\beta}^{ i_\beta j_\alpha}\\
A_{8} &=  (1 - \hat{n}(i_\alpha) - \hat{n}(j_\beta))\hat{G}_{a_\alpha b_\beta}^{ a_\beta b_\alpha}\\
A_{9} &=  (\hat{n}(i_\alpha) - \hat{n}(a_\alpha))\hat{G}_{j_\alpha b_\beta}^{ j_\beta b_\alpha}\\
A_{10} &=  (\hat{n}(b_\beta) - \hat{n}(j_\beta))\hat{G}_{i_\alpha a_\beta}^{ i_\beta a_\alpha}\\
A_{11} &=  -(1 - \hat{n}(i_\beta) - \hat{n}(j_\alpha))\hat{G}_{a_\alpha b_\beta}^{ a_\beta b_\alpha}\\
A_{12} &=  -(1 - \hat{n}(a_\beta) - \hat{n}(b_\alpha))\hat{G}_{i_\alpha j_\beta}^{ i_\beta j_\alpha}\\
A_{13} &=  (\hat{n}(j_\alpha) - \hat{n}(b_\alpha))\hat{G}_{i_\alpha a_\beta}^{ i_\beta a_\alpha}\\
A_{14} &=  (\hat{n}(a_\beta) - \hat{n}(i_\beta))\hat{G}_{j_\alpha b_\beta}^{ j_\beta b_\alpha}\\
A_{15} &=  (\hat{n}(j_\alpha) - \hat{n}(a_\alpha))\hat{G}_{i_\alpha b_\beta}^{ i_\beta b_\alpha}\\
A_{16} &=  (\hat{n}(b_\beta) - \hat{n}(i_\beta))\hat{G}_{j_\alpha a_\beta}^{ j_\beta a_\alpha}\\
A_{17} &=  (\hat{n}(i_\alpha) - \hat{n}(b_\alpha))\hat{G}_{j_\alpha a_\beta}^{ j_\beta a_\alpha}\\
A_{18} &=  (\hat{n}(a_\beta) - \hat{n}(j_\beta))\hat{G}_{i_\alpha b_\beta}^{ i_\beta b_\alpha}
\end{align*}
\newpage

{\it Second-Level Commutators:}
\begin{align*}
A_{19} &=  -(1 - \hat{n}(i_\alpha) - \hat{n}(j_\beta) + 2\hat{n}(i_\alpha)\hat{n}(j_\beta))\hat{G}_{i_\beta j_\alpha}^{ a_\alpha b_\beta}\\
A_{20} &=  -(1 - \hat{n}(a_\alpha) - \hat{n}(b_\beta) + 2\hat{n}(a_\alpha)\hat{n}(b_\beta))\hat{G}_{i_\alpha j_\beta}^{ a_\beta b_\alpha}\\
A_{21} &=  -(\hat{n}(j_\beta) + \hat{n}(b_\beta) - 2\hat{n}(j_\beta)\hat{n}(b_\beta))\hat{G}_{i_\alpha j_\alpha}^{ a_\alpha b_\alpha}\\
A_{22} &=  -(\hat{n}(i_\alpha) + \hat{n}(a_\alpha) - 2\hat{n}(i_\alpha)\hat{n}(a_\alpha))\hat{G}_{i_\beta j_\beta}^{ a_\beta b_\beta}\\
A_{23} &=  (1 - \hat{n}(i_\beta) - \hat{n}(j_\alpha) - \hat{n}(a_\alpha) - \hat{n}(b_\beta) + \hat{n}(i_\beta)\hat{n}(b_\beta)\\\nonumber
&\quad + \hat{n}(i_\beta)\hat{n}(a_\alpha) + \hat{n}(j_\alpha)\hat{n}(b_\beta) + \hat{n}(j_\alpha)\hat{n}(a_\alpha))\hat{G}_{i_\alpha j_\beta}^{ a_\beta b_\alpha}\\
A_{24} &=  (1 - \hat{n}(i_\alpha) - \hat{n}(j_\beta) - \hat{n}(a_\beta) - \hat{n}(b_\alpha) + \hat{n}(i_\alpha)\hat{n}(a_\beta) + \hat{n}(i_\alpha)\hat{n}(b_\alpha)\\\nonumber
&\quad + \hat{n}(j_\beta)\hat{n}(a_\beta) + \hat{n}(j_\beta)\hat{n}(b_\alpha))\hat{G}_{i_\beta j_\alpha}^{ a_\alpha b_\beta}\\
A_{25} &=  (\hat{n}(i_\alpha)\hat{n}(b_\alpha) + \hat{n}(j_\alpha)\hat{n}(a_\alpha) - \hat{n}(i_\alpha)\hat{n}(j_\alpha) - \hat{n}(a_\alpha)\hat{n}(b_\alpha))\hat{G}_{i_\beta j_\beta}^{ a_\beta b_\beta}\\
A_{26} &=  (\hat{n}(i_\beta)\hat{n}(b_\beta) + \hat{n}(j_\beta)\hat{n}(a_\beta) - \hat{n}(i_\beta)\hat{n}(j_\beta) - \hat{n}(a_\beta)\hat{n}(b_\beta))\hat{G}_{i_\alpha j_\alpha}^{ a_\alpha b_\alpha}\\
A_{27} &=  -(1 - \hat{n}(a_\beta) - \hat{n}(b_\alpha) + 2\hat{n}(a_\beta)\hat{n}(b_\alpha))\hat{G}_{i_\beta j_\alpha}^{ a_\alpha b_\beta}\\
A_{28} &=  -(1 - \hat{n}(i_\beta) - \hat{n}(j_\alpha) + 2\hat{n}(i_\beta)\hat{n}(j_\alpha))\hat{G}_{i_\alpha j_\beta}^{ a_\beta b_\alpha}\\
A_{29} &=  -(\hat{n}(i_\beta) + \hat{n}(a_\beta) - 2\hat{n}(i_\beta)\hat{n}(a_\beta))\hat{G}_{i_\alpha j_\alpha}^{ a_\alpha b_\alpha}\\
A_{30} &=  -(\hat{n}(j_\alpha) + \hat{n}(b_\alpha) - 2\hat{n}(j_\alpha)\hat{n}(b_\alpha))\hat{G}_{i_\beta j_\beta}^{ a_\beta b_\beta}\\
A_{31} &=  (1 - \hat{n}(i_\beta) - \hat{n}(j_\alpha) + 2\hat{n}(i_\beta)\hat{n}(j_\alpha))\hat{G}_{i_\alpha j_\beta}^{ a_\alpha b_\beta}\\
A_{32} &=  -(1 - \hat{n}(i_\alpha) - \hat{n}(j_\beta) - \hat{n}(a_\alpha) - \hat{n}(b_\beta) + \hat{n}(i_\alpha)\hat{n}(a_\alpha)\\\nonumber
&\quad + \hat{n}(i_\alpha)\hat{n}(b_\beta) + \hat{n}(j_\beta)\hat{n}(a_\alpha) + \hat{n}(j_\beta)\hat{n}(b_\beta))\hat{G}_{i_\beta j_\alpha}^{ a_\beta b_\alpha}\\
A_{33} &=  (1 - \hat{n}(a_\alpha) - \hat{n}(b_\beta) + 2\hat{n}(a_\alpha)\hat{n}(b_\beta))\hat{G}_{i_\beta j_\alpha}^{ a_\beta b_\alpha}\\
A_{34} &=  -(1 - \hat{n}(i_\beta) - \hat{n}(j_\alpha) - \hat{n}(a_\beta) - \hat{n}(b_\alpha) + \hat{n}(i_\beta)\hat{n}(a_\beta)\\\nonumber
&\quad + \hat{n}(i_\beta)\hat{n}(b_\alpha) + \hat{n}(j_\alpha)\hat{n}(a_\beta) + \hat{n}(j_\alpha)\hat{n}(b_\alpha))\hat{G}_{i_\alpha j_\beta}^{ a_\alpha b_\beta}\\
A_{35} &=  -(\hat{n}(i_\beta) + \hat{n}(b_\beta) - 2\hat{n}(i_\beta)\hat{n}(b_\beta))\hat{G}_{i_\alpha j_\alpha}^{ a_\alpha b_\alpha}\\
A_{36} &=  -(\hat{n}(j_\alpha) + \hat{n}(a_\alpha) - 2\hat{n}(j_\alpha)\hat{n}(a_\alpha))\hat{G}_{i_\beta j_\beta}^{ a_\beta b_\beta}\\
A_{37} &=  (\hat{n}(i_\alpha)\hat{n}(a_\alpha) + \hat{n}(j_\alpha)\hat{n}(b_\alpha) - \hat{n}(i_\alpha)\hat{n}(j_\alpha) - \hat{n}(a_\alpha)\hat{n}(b_\alpha))\hat{G}_{i_\beta j_\beta}^{ a_\beta b_\beta}\\
A_{38} &=  (\hat{n}(i_\beta)\hat{n}(a_\beta) + \hat{n}(j_\beta)\hat{n}(b_\beta) - \hat{n}(i_\beta)\hat{n}(j_\beta) - \hat{n}(a_\beta)\hat{n}(b_\beta))\hat{G}_{i_\alpha j_\alpha}^{ a_\alpha b_\alpha}\\
A_{39} &=  (1 - \hat{n}(a_\beta) - \hat{n}(b_\alpha) + 2\hat{n}(a_\beta)\hat{n}(b_\alpha))\hat{G}_{i_\alpha j_\beta}^{ a_\alpha b_\beta}
\end{align*}
\begin{align*}
A_{40} &=  (1 - \hat{n}(i_\alpha) - \hat{n}(j_\beta) + 2\hat{n}(i_\alpha)\hat{n}(j_\beta))\hat{G}_{i_\beta j_\alpha}^{ a_\beta b_\alpha}\\
A_{41} &=  (\hat{n}(j_\alpha) + \hat{n}(b_\alpha) - 2\hat{n}(j_\alpha)\hat{n}(b_\alpha))\hat{G}_{i_\alpha j_\beta}^{ a_\alpha b_\beta}\\
A_{42} &=  (\hat{n}(i_\alpha)\hat{n}(j_\beta) - \hat{n}(i_\alpha)\hat{n}(b_\beta) - \hat{n}(j_\beta)\hat{n}(a_\alpha) + \hat{n}(a_\alpha)\hat{n}(b_\beta))\hat{G}_{i_\beta j_\alpha}^{ a_\beta b_\alpha}\\
A_{43} &=  (\hat{n}(i_\alpha) + \hat{n}(a_\alpha) - 2\hat{n}(i_\alpha)\hat{n}(a_\alpha))\hat{G}_{i_\beta j_\alpha}^{ a_\beta b_\alpha}\\
A_{44} &=  (\hat{n}(i_\beta)\hat{n}(j_\alpha) - \hat{n}(i_\beta)\hat{n}(b_\alpha) - \hat{n}(j_\alpha)\hat{n}(a_\beta) + \hat{n}(a_\beta)\hat{n}(b_\alpha))\hat{G}_{i_\alpha j_\beta}^{ a_\alpha b_\beta}\\
A_{45} &=  (\hat{n}(i_\alpha) + \hat{n}(b_\alpha) - 2\hat{n}(i_\alpha)\hat{n}(b_\alpha))\hat{G}_{i_\beta j_\alpha}^{ a_\alpha b_\beta}\\
A_{46} &=  (\hat{n}(i_\beta)\hat{n}(j_\alpha) - \hat{n}(i_\beta)\hat{n}(a_\alpha) - \hat{n}(j_\alpha)\hat{n}(b_\beta) + \hat{n}(a_\alpha)\hat{n}(b_\beta))\hat{G}_{i_\alpha j_\beta}^{ a_\beta b_\alpha}\\
A_{47} &=  (\hat{n}(j_\alpha) + \hat{n}(a_\alpha) - 2\hat{n}(j_\alpha)\hat{n}(a_\alpha))\hat{G}_{i_\alpha j_\beta}^{ a_\beta b_\alpha}\\
A_{48} &=  (\hat{n}(i_\alpha)\hat{n}(j_\beta) - \hat{n}(i_\alpha)\hat{n}(a_\beta) - \hat{n}(j_\beta)\hat{n}(b_\alpha) + \hat{n}(a_\beta)\hat{n}(b_\alpha))\hat{G}_{i_\beta j_\alpha}^{ a_\alpha b_\beta}
\end{align*}

{\it Third-Level Commutators:}

\begin{align*}
A_{49} &=  (1 - \hat{n}(i_\alpha) - \hat{n}(i_\beta) - \hat{n}(j_\alpha) - \hat{n}(j_\beta) + \hat{n}(i_\alpha)\hat{n}(i_\beta) + \hat{n}(i_\alpha)\hat{n}(j_\alpha)+ \hat{n}(i_\beta)\hat{n}(j_\beta)\\\nonumber
&\quad + \hat{n}(j_\alpha)\hat{n}(j_\beta) + 2\hat{n}(i_\alpha)\hat{n}(j_\beta) - 2\hat{n}(i_\alpha)\hat{n}(i_\beta)\hat{n}(j_\beta) - 2\hat{n}(i_\alpha)\hat{n}(j_\alpha)\hat{n}(j_\beta))\hat{G}_{a_\alpha b_\beta}^{ a_\beta b_\alpha}\\
A_{50} &=  (1 - \hat{n}(a_\alpha) - \hat{n}(a_\beta) - \hat{n}(b_\alpha) - \hat{n}(b_\beta) + \hat{n}(a_\alpha)\hat{n}(a_\beta) + \hat{n}(a_\alpha)\hat{n}(b_\alpha) + \hat{n}(a_\beta)\hat{n}(b_\beta)\\\nonumber
&\quad + \hat{n}(b_\alpha)\hat{n}(b_\beta) + 2\hat{n}(a_\alpha)\hat{n}(b_\beta) - 2\hat{n}(a_\alpha)\hat{n}(a_\beta)\hat{n}(b_\beta) - 2\hat{n}(a_\alpha)\hat{n}(b_\alpha)\hat{n}(b_\beta))\hat{G}_{i_\alpha j_\beta}^{ i_\beta j_\alpha}\\
A_{51} &=  (\hat{n}(j_\beta)\hat{n}(b_\alpha) + \hat{n}(b_\alpha)\hat{n}(b_\beta) - \hat{n}(j_\alpha)\hat{n}(j_\beta) - \hat{n}(j_\alpha)\hat{n}(b_\beta)\\\nonumber
&\quad + 2\hat{n}(j_\alpha)\hat{n}(j_\beta)\hat{n}(b_\beta) - 2\hat{n}(j_\beta)\hat{n}(b_\alpha)\hat{n}(b_\beta))\hat{G}_{i_\alpha a_\beta}^{ i_\beta a_\alpha}\\
A_{52} &=  (\hat{n}(i_\alpha)\hat{n}(i_\beta) - \hat{n}(i_\alpha)\hat{n}(a_\beta) + \hat{n}(i_\beta)\hat{n}(a_\alpha) - \hat{n}(a_\alpha)\hat{n}(a_\beta)\\\nonumber
&\quad - 2\hat{n}(i_\alpha)\hat{n}(i_\beta)\hat{n}(a_\alpha) + 2\hat{n}(i_\alpha)\hat{n}(a_\alpha)\hat{n}(a_\beta))\hat{G}_{j_\alpha b_\beta}^{ j_\beta b_\alpha}\\
A_{53} &=  -(1 - \hat{n}(a_\alpha) - \hat{n}(a_\beta) - \hat{n}(b_\alpha) - \hat{n}(b_\beta) + \hat{n}(a_\alpha)\hat{n}(a_\beta) + \hat{n}(a_\alpha)\hat{n}(b_\alpha) + \hat{n}(a_\beta)\hat{n}(b_\beta)\\\nonumber
&\quad + \hat{n}(b_\alpha)\hat{n}(b_\beta) + 2\hat{n}(a_\beta)\hat{n}(b_\alpha)-  2\hat{n}(a_\alpha)\hat{n}(a_\beta)\hat{n}(b_\alpha) - 2\hat{n}(a_\beta)\hat{n}(b_\alpha)\hat{n}(b_\beta))\hat{G}_{i_\alpha j_\beta}^{ i_\beta j_\alpha}\\
A_{54} &=  -(1 - \hat{n}(i_\alpha) - \hat{n}(i_\beta) - \hat{n}(j_\alpha) - \hat{n}(j_\beta) + \hat{n}(i_\alpha)\hat{n}(i_\beta) + \hat{n}(i_\alpha)\hat{n}(j_\alpha) + \hat{n}(i_\beta)\hat{n}(j_\beta)\\\nonumber
&\quad + \hat{n}(j_\alpha)\hat{n}(j_\beta) + 2\hat{n}(i_\beta)\hat{n}(j_\alpha) - 2\hat{n}(i_\alpha)\hat{n}(i_\beta)\hat{n}(j_\alpha) - 2\hat{n}(i_\beta)\hat{n}(j_\alpha)\hat{n}(j_\beta))\hat{G}_{a_\alpha b_\beta}^{ a_\beta b_\alpha}\\
A_{55} &=  (\hat{n}(a_\alpha)\hat{n}(a_\beta) - \hat{n}(i_\alpha)\hat{n}(i_\beta) - \hat{n}(i_\alpha)\hat{n}(a_\beta) + \hat{n}(i_\beta)\hat{n}(a_\alpha)\\\nonumber
&\quad + 2\hat{n}(i_\alpha)\hat{n}(i_\beta)\hat{n}(a_\beta) - 2\hat{n}(i_\beta)\hat{n}(a_\alpha)\hat{n}(a_\beta))\hat{G}_{j_\alpha b_\beta}^{ j_\beta b_\alpha}
\end{align*}
\begin{align*}
A_{56} &=  (\hat{n}(j_\alpha)\hat{n}(j_\beta) - \hat{n}(j_\alpha)\hat{n}(b_\beta) + \hat{n}(j_\beta)\hat{n}(b_\alpha) - \hat{n}(b_\alpha)\hat{n}(b_\beta)\\\nonumber
&\quad - 2\hat{n}(j_\alpha)\hat{n}(j_\beta)\hat{n}(b_\alpha) + 2\hat{n}(j_\alpha)\hat{n}(b_\alpha)\hat{n}(b_\beta))\hat{G}_{i_\alpha a_\beta}^{ i_\beta a_\alpha}\\
A_{57} &=  -(\hat{n}(i_\alpha) - \hat{n}(i_\alpha)\hat{n}(i_\beta) - \hat{n}(i_\alpha)\hat{n}(a_\alpha) - \hat{n}(i_\beta)\hat{n}(a_\alpha) + 2\hat{n}(i_\alpha)\hat{n}(i_\beta)\hat{n}(a_\alpha))\hat{G}_{j_\alpha b_\beta}^{ j_\beta b_\alpha}\\
A_{58} &=  (\hat{n}(j_\beta) - \hat{n}(j_\alpha)\hat{n}(j_\beta) - \hat{n}(j_\alpha)\hat{n}(b_\beta) - \hat{n}(j_\beta)\hat{n}(b_\beta) + 2\hat{n}(j_\alpha)\hat{n}(j_\beta)\hat{n}(b_\beta))\hat{G}_{i_\alpha a_\beta}^{ i_\beta a_\alpha}\\
A_{59} &=  (\hat{n}(b_\alpha) - \hat{n}(a_\alpha)\hat{n}(b_\alpha) - \hat{n}(a_\alpha)\hat{n}(b_\beta) - \hat{n}(b_\alpha)\hat{n}(b_\beta) + 2\hat{n}(a_\alpha)\hat{n}(b_\alpha)\hat{n}(b_\beta))\hat{G}_{i_\alpha j_\beta}^{ i_\beta j_\alpha}\\
A_{60} &=  (\hat{n}(j_\alpha) - \hat{n}(i_\alpha)\hat{n}(j_\alpha) - \hat{n}(i_\alpha)\hat{n}(j_\beta) - \hat{n}(j_\alpha)\hat{n}(j_\beta) + 2\hat{n}(i_\alpha)\hat{n}(j_\alpha)\hat{n}(j_\beta))\hat{G}_{a_\alpha b_\beta}^{ a_\beta b_\alpha}\\
A_{61} &=  (2\hat{n}(b_\beta)\hat{n}(b_\alpha)\hat{n}(j_\alpha) - \hat{n}(b_\alpha)\hat{n}(j_\alpha) + \hat{n}(b_\alpha) - \hat{n}(b_\beta)\hat{n}(b_\alpha) - \hat{n}(b_\beta)\hat{n}(j_\alpha))\hat{G}_{i_\alpha a_\beta}^{ i_\beta a_\alpha}\\
A_{62} &=  (\hat{n}(a_\beta)\hat{n}(i_\beta) - \hat{n}(a_\beta) + \hat{n}(a_\beta)\hat{n}(a_\alpha) + \hat{n}(a_\alpha)\hat{n}(i_\beta) - 2\hat{n}(a_\beta)\hat{n}(a_\alpha)\hat{n}(i_\beta))\hat{G}_{j_\alpha b_\beta}^{ j_\beta b_\alpha}\\
A_{63} &=  (\hat{n}(j_\alpha)\hat{n}(i_\alpha) + \hat{n}(i_\beta)\hat{n}(i_\alpha) - \hat{n}(i_\alpha) + \hat{n}(j_\alpha)\hat{n}(i_\beta) - 2\hat{n}(j_\alpha)\hat{n}(i_\beta)\hat{n}(i_\alpha))\hat{G}_{a_\alpha b_\beta}^{ a_\beta b_\alpha}\\
A_{64} &=  (\hat{n}(b_\alpha)\hat{n}(a_\alpha) + \hat{n}(a_\beta)\hat{n}(a_\alpha) - \hat{n}(a_\alpha) - 2\hat{n}(b_\alpha)\hat{n}(a_\beta)\hat{n}(a_\alpha) + \hat{n}(b_\alpha)\hat{n}(a_\beta))\hat{G}_{i_\alpha j_\beta}^{ i_\beta j_\alpha}\\
A_{65} &=  (\hat{n}(j_\beta)\hat{n}(j_\alpha) + \hat{n}(a_\alpha)\hat{n}(j_\alpha) - \hat{n}(j_\alpha) - 2\hat{n}(a_\alpha)\hat{n}(j_\beta)\hat{n}(j_\alpha) + \hat{n}(a_\alpha)\hat{n}(j_\beta))\hat{G}_{i_\alpha b_\beta}^{ i_\beta b_\alpha}\\
A_{66} &=  (\hat{n}(i_\beta) - \hat{n}(i_\beta)\hat{n}(i_\alpha) - \hat{n}(b_\beta)\hat{n}(i_\beta) + 2\hat{n}(b_\beta)\hat{n}(i_\beta)\hat{n}(i_\alpha) - \hat{n}(b_\beta)\hat{n}(i_\alpha))\hat{G}_{j_\alpha a_\beta}^{ j_\beta a_\alpha}\\
A_{67} &=  (\hat{n}(b_\beta)\hat{n}(b_\alpha)\hat{n}(i_\alpha) - \hat{n}(b_\alpha)\hat{n}(i_\beta)\hat{n}(i_\alpha) - \hat{n}(b_\beta)\hat{n}(b_\alpha)\hat{n}(i_\beta) + \hat{n}(b_\beta)\hat{n}(i_\beta)\hat{n}(i_\alpha)\\\nonumber
&\quad + \hat{n}(b_\alpha)\hat{n}(i_\beta) - \hat{n}(b_\beta)\hat{n}(i_\alpha))\hat{G}_{j_\alpha a_\beta}^{ j_\beta a_\alpha}\\
A_{68} &=  -(\hat{n}(a_\alpha)\hat{n}(j_\beta)\hat{n}(j_\alpha) - \hat{n}(a_\beta)\hat{n}(a_\alpha)\hat{n}(j_\alpha) - \hat{n}(a_\beta)\hat{n}(j_\beta)\hat{n}(j_\alpha) + \hat{n}(a_\beta)\hat{n}(a_\alpha)\hat{n}(j_\beta)\\\nonumber
&\quad + \hat{n}(a_\beta)\hat{n}(j_\alpha) - \hat{n}(a_\alpha)\hat{n}(j_\beta))\hat{G}_{i_\alpha b_\beta}^{ i_\beta b_\alpha}\\
A_{69} &=  -(\hat{n}(a_\alpha)\hat{n}(j_\alpha) - \hat{n}(a_\alpha) - 2\hat{n}(a_\beta)\hat{n}(a_\alpha)\hat{n}(j_\alpha) + \hat{n}(a_\beta)\hat{n}(a_\alpha) + \hat{n}(a_\beta)\hat{n}(j_\alpha))\hat{G}_{i_\alpha b_\beta}^{ i_\beta b_\alpha}\\
A_{70} &=  (\hat{n}(b_\beta)\hat{n}(i_\beta) - \hat{n}(b_\beta) - 2\hat{n}(b_\beta)\hat{n}(b_\alpha)\hat{n}(i_\beta) + \hat{n}(b_\beta)\hat{n}(b_\alpha) + \hat{n}(b_\alpha)\hat{n}(i_\beta))\hat{G}_{j_\alpha a_\beta}^{ j_\beta a_\alpha}\\
A_{71} &=  (2\hat{n}(b_\beta)\hat{n}(b_\alpha)\hat{n}(i_\alpha) - \hat{n}(b_\alpha)\hat{n}(i_\alpha) + \hat{n}(b_\alpha) - \hat{n}(b_\beta)\hat{n}(b_\alpha) - \hat{n}(b_\beta)\hat{n}(i_\alpha))\hat{G}_{j_\alpha a_\beta}^{ j_\beta a_\alpha}\\
A_{72} &=  (\hat{n}(a_\beta)\hat{n}(j_\beta) - \hat{n}(a_\beta) + \hat{n}(a_\beta)\hat{n}(a_\alpha) + \hat{n}(a_\alpha)\hat{n}(j_\beta) - 2\hat{n}(a_\beta)\hat{n}(a_\alpha)\hat{n}(j_\beta))\hat{G}_{i_\alpha b_\beta}^{ i_\beta b_\alpha}\\
A_{73} &=  (1 - \hat{n}(i_\alpha) - \hat{n}(j_\beta) - \hat{n}(a_\beta) - \hat{n}(b_\alpha) + \hat{n}(i_\alpha)\hat{n}(b_\alpha) + \hat{n}(i_\alpha)\hat{n}(a_\beta) + \hat{n}(j_\beta)\hat{n}(b_\alpha)\\\nonumber
&\quad + \hat{n}(j_\beta)\hat{n}(a_\beta) + 2\hat{n}(i_\alpha)\hat{n}(j_\beta) + 2\hat{n}(a_\beta)\hat{n}(b_\alpha) - 2\hat{n}(i_\alpha)\hat{n}(a_\beta)\hat{n}(b_\alpha) - 2\hat{n}(j_\beta)\hat{n}(a_\beta)\hat{n}(b_\alpha)\\\nonumber
&\quad - 2\hat{n}(i_\alpha)\hat{n}(j_\beta)\hat{n}(b_\alpha) - 2\hat{n}(i_\alpha)\hat{n}(j_\beta)\hat{n}(a_\beta) + 4\hat{n}(i_\alpha)\hat{n}(j_\beta)\hat{n}(a_\beta)\hat{n}(b_\alpha))\hat{G}_{i_\beta j_\alpha}^{ a_\alpha b_\beta}\\
A_{74} &=  (1 - \hat{n}(i_\beta) - \hat{n}(j_\alpha) - \hat{n}(a_\alpha) - \hat{n}(b_\beta) + \hat{n}(j_\alpha)\hat{n}(a_\alpha) + \hat{n}(j_\alpha)\hat{n}(b_\beta) + \hat{n}(i_\beta)\hat{n}(a_\alpha)\\\nonumber
&\quad + \hat{n}(i_\beta)\hat{n}(b_\beta) + 2\hat{n}(i_\beta)\hat{n}(j_\alpha) + 2\hat{n}(a_\alpha)\hat{n}(b_\beta) - 2\hat{n}(i_\beta)\hat{n}(j_\alpha)\hat{n}(a_\alpha) - 2\hat{n}(i_\beta)\hat{n}(j_\alpha)\hat{n}(b_\beta)\\\nonumber
&\quad - 2\hat{n}(j_\alpha)\hat{n}(a_\alpha)\hat{n}(b_\beta) - 2\hat{n}(i_\beta)\hat{n}(a_\alpha)\hat{n}(b_\beta) + 4\hat{n}(i_\beta)\hat{n}(j_\alpha)\hat{n}(a_\alpha)\hat{n}(b_\beta))\hat{G}_{i_\alpha j_\beta}^{ a_\beta b_\alpha}
\end{align*}
\begin{align*}
A_{75} &=  (\hat{n}(a_\beta)\hat{n}(j_\beta) - 2\hat{n}(b_\beta)\hat{n}(j_\beta)\hat{n}(i_\beta) - 2\hat{n}(b_\beta)\hat{n}(a_\beta)\hat{n}(j_\beta) + \hat{n}(j_\beta)\hat{n}(i_\beta)\\\nonumber
&\quad + 4\hat{n}(b_\beta)\hat{n}(a_\beta)\hat{n}(j_\beta)\hat{n}(i_\beta) - 2\hat{n}(a_\beta)\hat{n}(j_\beta)\hat{n}(i_\beta) + \hat{n}(b_\beta)\hat{n}(i_\beta) + \hat{n}(b_\beta)\hat{n}(a_\beta)\\\nonumber
&\quad - 2\hat{n}(b_\beta)\hat{n}(a_\beta)\hat{n}(i_\beta))\hat{G}_{i_\alpha j_\alpha}^{ a_\alpha b_\alpha}\\
A_{76} &=  (\hat{n}(j_\alpha)\hat{n}(i_\alpha) + \hat{n}(b_\alpha)\hat{n}(i_\alpha) + 4\hat{n}(b_\alpha)\hat{n}(a_\alpha)\hat{n}(j_\alpha)\hat{n}(i_\alpha) - 2\hat{n}(b_\alpha)\hat{n}(j_\alpha)\hat{n}(i_\alpha)\\\nonumber
&\quad - 2\hat{n}(a_\alpha)\hat{n}(j_\alpha)\hat{n}(i_\alpha) - 2\hat{n}(b_\alpha)\hat{n}(a_\alpha)\hat{n}(i_\alpha) + \hat{n}(a_\alpha)\hat{n}(j_\alpha) + \hat{n}(b_\alpha)\hat{n}(a_\alpha)\\\nonumber
&\quad - 2\hat{n}(b_\alpha)\hat{n}(a_\alpha)\hat{n}(j_\alpha))\hat{G}_{i_\beta j_\beta}^{ a_\beta b_\beta}\\
A_{77} &=  (\hat{n}(j_\beta)\hat{n}(i_\beta) - \hat{n}(b_\beta)\hat{n}(i_\beta) - \hat{n}(b_\beta)\hat{n}(j_\beta) + \hat{n}(b_\beta))\hat{G}_{i_\alpha j_\alpha}^{ a_\alpha b_\alpha}\\
A_{78} &=  (\hat{n}(j_\alpha)\hat{n}(i_\alpha) - \hat{n}(a_\alpha)\hat{n}(j_\alpha) - \hat{n}(a_\alpha)\hat{n}(i_\alpha) + \hat{n}(a_\alpha))\hat{G}_{i_\beta j_\beta}^{ a_\beta b_\beta}\\
A_{79} &=  (\hat{n}(b_\alpha)\hat{n}(j_\beta) - \hat{n}(j_\beta)\hat{n}(i_\alpha) + \hat{n}(b_\alpha)\hat{n}(i_\alpha) - \hat{n}(b_\alpha))\hat{G}_{i_\beta j_\alpha}^{ a_\alpha b_\beta}\\
A_{80} &=  (\hat{n}(b_\beta)\hat{n}(j_\alpha) + \hat{n}(a_\alpha)\hat{n}(j_\alpha) - \hat{n}(j_\alpha) - \hat{n}(b_\beta)\hat{n}(a_\alpha))\hat{G}_{i_\alpha j_\beta}^{ a_\beta b_\alpha}\\
A_{81} &=  -(1 - \hat{n}(i_\alpha) - \hat{n}(j_\beta) - \hat{n}(a_\alpha) - \hat{n}(b_\beta) + \hat{n}(i_\alpha)\hat{n}(a_\alpha) + \hat{n}(i_\alpha)\hat{n}(b_\beta) + \hat{n}(j_\beta)\hat{n}(a_\alpha)\\\nonumber
&\quad + \hat{n}(j_\beta)\hat{n}(b_\beta) + 2\hat{n}(i_\alpha)\hat{n}(j_\beta) + 2\hat{n}(a_\alpha)\hat{n}(b_\beta) - 2\hat{n}(i_\alpha)\hat{n}(a_\alpha)\hat{n}(b_\beta) - 2\hat{n}(j_\beta)\hat{n}(a_\alpha)\hat{n}(b_\beta)\\\nonumber
&\quad - 2\hat{n}(i_\alpha)\hat{n}(j_\beta)\hat{n}(a_\alpha) - 2\hat{n}(i_\alpha)\hat{n}(j_\beta)\hat{n}(b_\beta) + 4\hat{n}(i_\alpha)\hat{n}(j_\beta)\hat{n}(a_\alpha)\hat{n}(b_\beta))\hat{G}_{i_\beta j_\alpha}^{ a_\beta b_\alpha}\\
A_{82} &=  -(1 - \hat{n}(i_\beta) - \hat{n}(j_\alpha) - \hat{n}(a_\beta) - \hat{n}(b_\alpha) + \hat{n}(i_\beta)\hat{n}(a_\beta) + \hat{n}(i_\beta)\hat{n}(b_\alpha) + \hat{n}(j_\alpha)\hat{n}(a_\beta)\\\nonumber
&\quad + \hat{n}(j_\alpha)\hat{n}(b_\alpha) + 2\hat{n}(i_\beta)\hat{n}(j_\alpha) + 2\hat{n}(a_\beta)\hat{n}(b_\alpha) - 2\hat{n}(i_\beta)\hat{n}(a_\beta)\hat{n}(b_\alpha) - 2\hat{n}(j_\alpha)\hat{n}(a_\beta)\hat{n}(b_\alpha)\\\nonumber
&\quad - 2\hat{n}(i_\beta)\hat{n}(j_\alpha)\hat{n}(a_\beta) - 2\hat{n}(i_\beta)\hat{n}(j_\alpha)\hat{n}(b_\alpha) + 4\hat{n}(i_\beta)\hat{n}(j_\alpha)\hat{n}(a_\beta)\hat{n}(b_\alpha))\hat{G}_{i_\alpha j_\beta}^{ a_\alpha b_\beta}\\
A_{83} &=  (\hat{n}(b_\alpha) + \hat{n}(j_\alpha)\hat{n}(i_\beta) - \hat{n}(b_\alpha)\hat{n}(j_\alpha) - \hat{n}(b_\alpha)\hat{n}(i_\beta))\hat{G}_{i_\alpha j_\beta}^{ a_\alpha b_\beta}\\
A_{84} &=  (\hat{n}(b_\beta)\hat{n}(a_\alpha) - \hat{n}(b_\beta)\hat{n}(i_\alpha) - \hat{n}(a_\alpha)\hat{n}(i_\alpha) + \hat{n}(i_\alpha))\hat{G}_{i_\beta j_\alpha}^{ a_\beta b_\alpha}
\end{align*}

\subsection*{Commuting groups}

The groups of commuting generators are,

\begin{align}
s&=\{1,2,15,16,17,18,31,32,33,34,39,40,41,42,43,44,65,66,67,68,69,70,71,72,81,82,83,84\}\\
s&=\{3,4,9,10,13,14,19,20,23,24,27,28,45,46,47,48,51,52,55,56,57,58,61,62,73,74,79,80\}\\
s&=\{5,6,7,8,11,12,21,22,25,26,29,30,35,36,37,38,49,50,53,54,59,60,63,64,75,76,77,78\}
\end{align}
 
\subsection*{Algebraic structure}

%The Killing form has rank $132$ and nullity $6$.
Let us denote $\mathfrak s=[\mathcal A_{S4T},\mathcal A_{S4T}]$ a derived algebra that constitute the semi-simple part,  $\dim \mathfrak s=78$. There is a 6 dimensional center, $Z(\mathcal A_{S4T})$. 
The solvable radical coincides with the center: $\mathfrak r = Z(\mathcal A_{S4T})$, and
\[ \mathcal A_{S4T} \;\cong\; \mathfrak s \;\oplus\; Z(\mathcal A_{S4T}) \]
is a {\it reductive} Lie algebra (direct sum; $[\mathfrak s, Z(\mathcal A_{S4T})] = 0$).

%section*{Structure of the $84$-generator fermionic Lie algebra}

The algebra $\mathcal A_{S4T}$ is reductive with
\[\dim\mathcal A_{S4T}=84,\qquad \dim Z(\mathcal A_{S4T})=6,\qquad \dim[\mathcal A_{S4T},\mathcal A_{S4T}]=78.\]
It decomposes as a direct sum of ideals
\[\mathcal A_{S4T} \;=\; Z(\mathcal A_{S4T})\;\oplus\;\bigoplus_{r=1}^{26}\mathfrak s_r,\]
where each $\mathfrak s_r$ is a $3$-dimensional simple ideal (isomorphic to $\mathfrak{sl}_2$ over $\mathbb C$), and $[\mathfrak s_r,\mathfrak s_{r'}]=0$ for $r\neq r'$.
\subsection*{Center $Z(\mathcal A_{S4T})$}
A basis $\{Z_1,\dots,Z_6\}$ of the center is:
\begin{align*}
Z_1 &= 2A_{3} + 2A_{19} + A_{24} + 2A_{27} + 2A_{48} + A_{73}\\
Z_2 &= 2A_{4} + 2A_{20} + A_{23} + 2A_{28} + 2A_{46} + A_{74}\\
Z_3 &= 2A_{5} + 2A_{21} + A_{26} + 2A_{29} - 2A_{38} + A_{75}\\
Z_4 &= 2A_{6} + 2A_{22} + A_{25} + 2A_{30} - 2A_{37} + A_{76}\\
Z_5 &= -2A_{2} + A_{32} + 2A_{33} + 2A_{40} - 2A_{42} + A_{81}\\
Z_6 &= -2A_{1} + 2A_{31} + A_{34} + 2A_{39} - 2A_{44} + A_{82}\\
\end{align*}
\subsection*{Semisimple part $[\mathcal A_{S4T},\mathcal A_{S4T}]$}
We list the $26$ commuting simple ideals $\mathfrak s_r$ (each of dimension $3$) via generators
\[\mathfrak s_r=\mathrm{span}\{B^{(r)}_1,B^{(r)}_2,B^{(r)}_3\},\qquad B^{(r)}_m=\sum_k c^{(r,m)}_k A_k.\]
\paragraph{Ideal $\mathfrak s_{1}$ (dim $3$).}
\begin{align*}
B^{(1)}_1 &= -A_{32} - A_{34} + 2A_{67} + 2A_{68} + A_{81} + A_{82}\\
B^{(1)}_2 &= -A_{25} - A_{26} - A_{49} - A_{50} - A_{53} - A_{54} + A_{75} + A_{76}\\
B^{(1)}_3 &= -A_{23} - A_{24} + A_{51} + A_{52} + A_{55} + A_{56} + A_{73} + A_{74}\\
\end{align*}
\paragraph{Ideal $\mathfrak s_{2}$ (dim $3$).}
\begin{align*}
B^{(2)}_1 &= -A_{32} - A_{34} - 2A_{67} - 2A_{68} + A_{81} + A_{82}\\
B^{(2)}_2 &= -A_{25} - A_{26} + A_{49} + A_{50} + A_{53} + A_{54} + A_{75} + A_{76}\\
B^{(2)}_3 &= -A_{23} - A_{24} - A_{51} - A_{52} - A_{55} - A_{56} + A_{73} + A_{74}\\
\end{align*}
\paragraph{Ideal $\mathfrak s_{3}$ (dim $3$).}
\begin{align*}
B^{(3)}_1 &= A_{32} - A_{34} + 2A_{67} - 2A_{68} - A_{81} + A_{82}\\
B^{(3)}_2 &= -A_{25} + A_{26} + A_{49} - A_{50} - A_{53} + A_{54} - A_{75} + A_{76}\\
B^{(3)}_3 &= -A_{23} + A_{24} - A_{51} + A_{52} + A_{55} - A_{56} - A_{73} + A_{74}\\
\end{align*}
\paragraph{Ideal $\mathfrak s_{4}$ (dim $3$).}
\begin{align*}
B^{(4)}_1 &= A_{32} - A_{34} - 2A_{67} + 2A_{68} - A_{81} + A_{82}\\
B^{(4)}_2 &= -A_{25} + A_{26} - A_{49} + A_{50} + A_{53} - A_{54} - A_{75} + A_{76}\\
B^{(4)}_3 &= -A_{23} + A_{24} + A_{51} - A_{52} - A_{55} + A_{56} - A_{73} + A_{74}\\
\end{align*}
\paragraph{Ideal $\mathfrak s_{5}$ (dim $3$).}
\begin{align*}
B^{(5)}_1 &= -\frac{1}{2}A_{34} - A_{65} + A_{68} + \frac{1}{2}A_{82} + A_{83}\\
B^{(5)}_2 &= -\frac{1}{2}A_{24} + \frac{1}{2}A_{52} + \frac{1}{2}A_{55} - A_{57} + \frac{1}{2}A_{73} + A_{79}\\
B^{(5)}_3 &= \frac{1}{2}A_{26} + \frac{1}{2}A_{50} + \frac{1}{2}A_{53} + A_{59} - \frac{1}{2}A_{75} + A_{77}\\
\end{align*}
\paragraph{Ideal $\mathfrak s_{6}$ (dim $3$).}
\begin{align*}
B^{(6)}_1 &= -\frac{1}{2}A_{34} + A_{65} - A_{68} + \frac{1}{2}A_{82} + A_{83}\\
B^{(6)}_2 &= -\frac{1}{2}A_{24} - \frac{1}{2}A_{52} - \frac{1}{2}A_{55} + A_{57} + \frac{1}{2}A_{73} + A_{79}\\
B^{(6)}_3 &= \frac{1}{2}A_{26} - \frac{1}{2}A_{50} - \frac{1}{2}A_{53} - A_{59} - \frac{1}{2}A_{75} + A_{77}\\
\end{align*}
\paragraph{Ideal $\mathfrak s_{7}$ (dim $3$).}
\begin{align*}
B^{(7)}_1 &= -A_{31} - \frac{1}{2}A_{34} - A_{66} + A_{67} - \frac{1}{2}A_{82} + A_{83}\\
B^{(7)}_2 &= A_{7} - A_{19} - \frac{1}{2}A_{24} - \frac{1}{2}A_{25} - \frac{1}{2}A_{50} + \frac{1}{2}A_{51} + \frac{1}{2}A_{53} + \frac{1}{2}A_{56} - A_{58} - A_{59} - \frac{1}{2}A_{73} + \frac{1}{2}A_{76} - A_{78} + A_{79}\\
B^{(7)}_3 &= -A_{7} + \frac{1}{2}A_{25} + \frac{1}{2}A_{50} - \frac{1}{2}A_{53} + A_{59} - \frac{1}{2}A_{76} + A_{78}\\
\end{align*}
\paragraph{Ideal $\mathfrak s_{8}$ (dim $3$).}
\begin{align*}
B^{(8)}_1 &= -A_{31} - \frac{1}{2}A_{34} + A_{66} - A_{67} - \frac{1}{2}A_{82} + A_{83}\\
B^{(8)}_2 &= -A_{19} - \frac{1}{2}A_{24} - \frac{1}{2}A_{51} - \frac{1}{2}A_{56} + A_{58} - \frac{1}{2}A_{73} + A_{79}\\
B^{(8)}_3 &= A_{7} + \frac{1}{2}A_{25} - \frac{1}{2}A_{50} + \frac{1}{2}A_{53} - A_{59} - \frac{1}{2}A_{76} + A_{78}\\
\end{align*}
\paragraph{Ideal $\mathfrak s_{9}$ (dim $3$).}
\begin{align*}
B^{(9)}_1 &= A_{32} + A_{34} + A_{81} + A_{82}\\
B^{(9)}_2 &= A_{23} - A_{24} - A_{73} + A_{74}\\
B^{(9)}_3 &= -A_{49} + A_{50} - A_{53} + A_{54}\\
\end{align*}
\paragraph{Ideal $\mathfrak s_{10}$ (dim $3$).}
\begin{align*}
B^{(10)}_1 &= -A_{32} + A_{34} - A_{81} + A_{82}\\
B^{(10)}_2 &= A_{23} + A_{24} + A_{73} + A_{74}\\
B^{(10)}_3 &= -A_{49} - A_{50} + A_{53} + A_{54}\\
\end{align*}
\paragraph{Ideal $\mathfrak s_{11}$ (dim $3$).}
\begin{align*}
B^{(11)}_1 &= \frac{1}{2}A_{34} - A_{41} - A_{44} + A_{67} - A_{71} - \frac{1}{2}A_{82} + A_{83}\\
B^{(11)}_2 &= -A_{9} - \frac{1}{2}A_{23} + \frac{1}{2}A_{52} - \frac{1}{2}A_{55} - A_{57} + \frac{1}{2}A_{74} + A_{80}\\
B^{(11)}_3 &= A_{21} + \frac{1}{2}A_{26} + \frac{1}{2}A_{49} + \frac{1}{2}A_{54} + A_{60} + \frac{1}{2}A_{75} + A_{77}\\
\end{align*}
\paragraph{Ideal $\mathfrak s_{12}$ (dim $3$).}
\begin{align*}
B^{(12)}_1 &= \frac{1}{2}A_{34} - A_{41} - A_{44} - A_{67} + A_{71} - \frac{1}{2}A_{82} + A_{83}\\
B^{(12)}_2 &= A_{9} - \frac{1}{2}A_{23} - \frac{1}{2}A_{52} + \frac{1}{2}A_{55} + A_{57} + \frac{1}{2}A_{74} + A_{80}\\
B^{(12)}_3 &= A_{21} + \frac{1}{2}A_{26} - \frac{1}{2}A_{49} - \frac{1}{2}A_{54} - A_{60} + \frac{1}{2}A_{75} + A_{77}\\
\end{align*}
\paragraph{Ideal $\mathfrak s_{13}$ (dim $3$).}
\begin{align*}
B^{(13)}_1 &= \frac{1}{2}A_{34} + A_{39} - A_{41} - A_{44} + A_{68} - A_{72} + \frac{1}{2}A_{82} + A_{83}\\
B^{(13)}_2 &= -A_{10} - A_{20} - \frac{1}{2}A_{23} + \frac{1}{2}A_{51} - \frac{1}{2}A_{56} - A_{58} - \frac{1}{2}A_{74} + A_{80}\\
B^{(13)}_3 &= -A_{8} + A_{22} + \frac{1}{2}A_{25} + \frac{1}{2}A_{49} - \frac{1}{2}A_{54} + A_{60} + \frac{1}{2}A_{76} + A_{78}\\
\end{align*}
\paragraph{Ideal $\mathfrak s_{14}$ (dim $3$).}
\begin{align*}
B^{(14)}_1 &= \frac{1}{2}A_{34} + A_{39} - A_{41} - A_{44} - A_{68} + A_{72} + \frac{1}{2}A_{82} + A_{83}\\
B^{(14)}_2 &= A_{10} - A_{20} - \frac{1}{2}A_{23} - \frac{1}{2}A_{51} + \frac{1}{2}A_{56} + A_{58} - \frac{1}{2}A_{74} + A_{80}\\
B^{(14)}_3 &= A_{8} + A_{22} + \frac{1}{2}A_{25} - \frac{1}{2}A_{49} + \frac{1}{2}A_{54} - A_{60} + \frac{1}{2}A_{76} + A_{78}\\
\end{align*}
\paragraph{Ideal $\mathfrak s_{15}$ (dim $3$).}
\begin{align*}
B^{(15)}_1 &= -A_{16} - A_{17} + \frac{1}{2}A_{32} - A_{42} - A_{43} - A_{66} + A_{67} - A_{70} - A_{71} - \frac{1}{2}A_{81} + A_{84}\\
B^{(15)}_2 &= -A_{13} + \frac{1}{2}A_{23} + \frac{1}{2}A_{26} + A_{29} - A_{35} - A_{38} + A_{46} + A_{47} - \frac{1}{2}A_{50} - \frac{1}{2}A_{51} - \frac{1}{2}A_{53} + \frac{1}{2}A_{56} - A_{61} - A_{64} - \frac{1}{2}A_{74} + \frac{1}{2}A_{75} - A_{77} + A_{80}\\
B^{(15)}_3 &= -\frac{1}{2}A_{26} - A_{29} + A_{35} + A_{38} + \frac{1}{2}A_{50} + \frac{1}{2}A_{53} + A_{64} - \frac{1}{2}A_{75} + A_{77}\\
\end{align*}
\paragraph{Ideal $\mathfrak s_{16}$ (dim $3$).}
\begin{align*}
B^{(16)}_1 &= A_{16} + A_{17} + \frac{1}{2}A_{32} - A_{42} - A_{43} + A_{66} - A_{67} + A_{70} + A_{71} - \frac{1}{2}A_{81} + A_{84}\\
B^{(16)}_2 &= A_{13} + \frac{1}{2}A_{23} + A_{46} + A_{47} + \frac{1}{2}A_{51} - \frac{1}{2}A_{56} + A_{61} - \frac{1}{2}A_{74} + A_{80}\\
B^{(16)}_3 &= -\frac{1}{2}A_{26} - A_{29} + A_{35} + A_{38} - \frac{1}{2}A_{50} - \frac{1}{2}A_{53} - A_{64} - \frac{1}{2}A_{75} + A_{77}\\
\end{align*}
\paragraph{Ideal $\mathfrak s_{17}$ (dim $3$).}
\begin{align*}
B^{(17)}_1 &= -A_{15} - A_{18} + \frac{1}{2}A_{32} + A_{40} - A_{42} - A_{43} - A_{65} + A_{68} - A_{69} - A_{72} + \frac{1}{2}A_{81} + A_{84}\\
B^{(17)}_2 &= -A_{14} + \frac{1}{2}A_{23} + A_{28} + A_{46} + A_{47} - \frac{1}{2}A_{52} + \frac{1}{2}A_{55} - A_{62} + \frac{1}{2}A_{74} + A_{80}\\
B^{(17)}_3 &= -A_{12} - \frac{1}{2}A_{25} - A_{30} + A_{36} + A_{37} - \frac{1}{2}A_{50} + \frac{1}{2}A_{53} + A_{64} - \frac{1}{2}A_{76} + A_{78}\\
\end{align*}
\paragraph{Ideal $\mathfrak s_{18}$ (dim $3$).}
\begin{align*}
B^{(18)}_1 &= A_{15} + A_{18} + \frac{1}{2}A_{32} + A_{40} - A_{42} - A_{43} + A_{65} - A_{68} + A_{69} + A_{72} + \frac{1}{2}A_{81} + A_{84}\\
B^{(18)}_2 &= A_{14} + \frac{1}{2}A_{23} + A_{28} + A_{46} + A_{47} + \frac{1}{2}A_{52} - \frac{1}{2}A_{55} + A_{62} + \frac{1}{2}A_{74} + A_{80}\\
B^{(18)}_3 &= A_{12} - \frac{1}{2}A_{25} - A_{30} + A_{36} + A_{37} + \frac{1}{2}A_{50} - \frac{1}{2}A_{53} - A_{64} - \frac{1}{2}A_{76} + A_{78}\\
\end{align*}
\paragraph{Ideal $\mathfrak s_{19}$ (dim $3$).}
\begin{align*}
B^{(19)}_1 &= -A_{25} + A_{26} - A_{32} - A_{34} + 2A_{42} + 2A_{44} + A_{75} - A_{76} + A_{81} + A_{82}\\
B^{(19)}_2 &= A_{25} - A_{26} - A_{75} + A_{76}\\
B^{(19)}_3 &= -A_{51} + A_{52} - A_{55} + A_{56}\\
\end{align*}
\paragraph{Ideal $\mathfrak s_{20}$ (dim $3$).}
\begin{align*}
B^{(20)}_1 &= A_{32} - A_{34} - 2A_{42} + 2A_{44} - A_{81} + A_{82}\\
B^{(20)}_2 &= A_{25} + A_{26} + A_{75} + A_{76}\\
B^{(20)}_3 &= -A_{51} - A_{52} + A_{55} + A_{56}\\
\end{align*}
\paragraph{Ideal $\mathfrak s_{21}$ (dim $3$).}
\begin{align*}
B^{(21)}_1 &= -A_{25} + A_{26} + 2A_{37} - 2A_{38} - A_{75} + A_{76}\\
B^{(21)}_2 &= -A_{23} - A_{24} - 2A_{46} - 2A_{48} + A_{73} + A_{74}\\
B^{(21)}_3 &= -A_{15} + A_{16} - A_{65} + A_{66} - A_{67} + A_{68} - A_{69} + A_{70}\\
\end{align*}
\paragraph{Ideal $\mathfrak s_{22}$ (dim $3$).}
\begin{align*}
B^{(22)}_1 &= -A_{25} - A_{26} + 2A_{37} + 2A_{38} + A_{75} + A_{76}\\
B^{(22)}_2 &= -A_{23} + A_{24} - 2A_{46} + 2A_{48} - A_{73} + A_{74}\\
B^{(22)}_3 &= A_{15} + A_{16} + A_{65} + A_{66} - A_{67} - A_{68} + A_{69} + A_{70}\\
\end{align*}
\paragraph{Ideal $\mathfrak s_{23}$ (dim $3$).}
\begin{align*}
B^{(23)}_1 &= -\frac{1}{2}A_{32} + A_{68} - A_{69} + \frac{1}{2}A_{81} + A_{84}\\
B^{(23)}_2 &= \frac{1}{2}A_{24} + A_{45} + A_{48} - \frac{1}{2}A_{51} - \frac{1}{2}A_{56} + A_{61} - \frac{1}{2}A_{73} + A_{79}\\
B^{(23)}_3 &= -\frac{1}{2}A_{26} + A_{35} + A_{38} - \frac{1}{2}A_{49} - \frac{1}{2}A_{54} - A_{63} + \frac{1}{2}A_{75} + A_{77}\\
\end{align*}
\paragraph{Ideal $\mathfrak s_{24}$ (dim $3$).}
\begin{align*}
B^{(24)}_1 &= -\frac{1}{2}A_{32} - A_{68} + A_{69} + \frac{1}{2}A_{81} + A_{84}\\
B^{(24)}_2 &= \frac{1}{2}A_{24} + A_{45} + A_{48} + \frac{1}{2}A_{51} + \frac{1}{2}A_{56} - A_{61} - \frac{1}{2}A_{73} + A_{79}\\
B^{(24)}_3 &= -\frac{1}{2}A_{26} + A_{35} + A_{38} + \frac{1}{2}A_{49} + \frac{1}{2}A_{54} + A_{63} + \frac{1}{2}A_{75} + A_{77}\\
\end{align*}
\paragraph{Ideal $\mathfrak s_{25}$ (dim $3$).}
\begin{align*}
B^{(25)}_1 &= -\frac{1}{2}A_{32} - A_{33} + A_{67} - A_{70} - \frac{1}{2}A_{81} + A_{84}\\
B^{(25)}_2 &= \frac{1}{2}A_{24} + A_{27} + A_{45} + A_{48} - \frac{1}{2}A_{52} - \frac{1}{2}A_{55} + A_{62} + \frac{1}{2}A_{73} + A_{79}\\
B^{(25)}_3 &= A_{11} - \frac{1}{2}A_{25} + A_{36} + A_{37} + \frac{1}{2}A_{49} - \frac{1}{2}A_{54} - A_{63} + \frac{1}{2}A_{76} + A_{78}\\
\end{align*}
\paragraph{Ideal $\mathfrak s_{26}$ (dim $3$).}
\begin{align*}
B^{(26)}_1 &= -\frac{1}{2}A_{32} - A_{33} - A_{67} + A_{70} - \frac{1}{2}A_{81} + A_{84}\\
B^{(26)}_2 &= \frac{1}{2}A_{24} + A_{27} + A_{45} + A_{48} + \frac{1}{2}A_{52} + \frac{1}{2}A_{55} - A_{62} + \frac{1}{2}A_{73} + A_{79}\\
B^{(26)}_3 &= -A_{11} - \frac{1}{2}A_{25} + A_{36} + A_{37} - \frac{1}{2}A_{49} + \frac{1}{2}A_{54} + A_{63} + \frac{1}{2}A_{76} + A_{78}\\
\end{align*}

\end{document}